\documentclass[12pt]{article} 

\usepackage[utf8]{inputenc}
\usepackage[english]{babel}

\usepackage{amsmath}
\usepackage{amsthm}
\usepackage{amssymb}
\usepackage[left=2.5cm,right=2.5cm,top=3cm,bottom=3cm,bindingoffset=0cm]{geometry}
\usepackage{dsfont}
\usepackage{bm}
\usepackage{leftidx}
\usepackage[makeroom]{cancel}
\usepackage{indentfirst}
\usepackage{caption}
\usepackage{hyperref}
\usepackage{tikz}
\usepackage{subfig}
\usepackage{graphicx}
\usepackage{tikz-cd}
\usepackage[titletoc,title]{appendix}

\usepackage{adjustbox}
\usepackage{mathtools}
\usepackage{multirow}
\usetikzlibrary{matrix,arrows,decorations.pathmorphing}
\newenvironment{sloppypar*}{\sloppy\ignorespaces}{\par}

\DeclareMathOperator{\sn}{sn\s{}  }
\DeclareMathOperator{\cn}{cn\s{}  }
\DeclareMathOperator{\dn}{dn\s{}  }
\DeclareMathOperator{\snt}{sn^2\s{}  }

\DeclareMathOperator{\cnt}{cn^2\s{}  }
\DeclareMathOperator{\cntt}{cn^4\s{}  }

\DeclareMathOperator{\Res}{Res\s{}  }

\newcommand{\bma} {\begin{pmatrix}}
\newcommand{\ema} {\end{pmatrix}}

\newcommand{\iu}{{i\mkern1mu}}

\renewcommand{\d}[1]{\ensuremath{\operatorname{d}\!{#1}}}
\newcommand{\me}{\mathrm{e}}
\newtheoremstyle{dotless}{}{}{\itshape}{}{\bfseries}{}{ }{}
\theoremstyle{dotless}

\newcommand{\qmE} {\varrho}
\newcommand{\qactE} {\varsigma}
\newcommand{\qm} {\rho}
\newcommand{\qact} {\sigma}

\newcommand{\lP} {\mathcal{P}}
\newcommand{\lQ} {\mathcal{Q}}
\newcommand{\s}{\hspace{.08em}}
\newcommand{\integrand}{\varphi}
\newcommand{\integrandform}{\lambda}
\newcommand{\exampleform}{\mu}

\let\oldFootnote\footnote
\newcommand\nextToken\relax
\renewcommand\footnote[1]{\oldFootnote{#1}\futurelet\nextToken\isFootnote}
\newcommand\isFootnote{\ifx\footnote\nextToken\textsuperscript{,}\fi}

\usepackage[nottoc]{tocbibind}

\usepackage[shortcuts]{extdash}

\usepackage{authblk}
\makeatletter
\renewcommand\AB@authnote[1]{\rlap{\textsuperscript{\normalfont#1}}}

\makeatother
\author[1,2]{Michael Kreshchuk\thanks{michael.kreshchuk @ tufts.edu}\s}
\author[3,2]{Tobias Gulden\thanks{gulden @ physics.technion.ac.il}\s}
\affil[1]{{\small
~Department of Physics, Tufts University, Medford, MA, 02155, USA
}}
\affil[2]{{\small
~School of Physics and Astronomy, University of Minnesota, Minneapolis, MN, 55455, USA
}}
\affil[3]{{\small
~Department of Physics, Technion~--- Israel Institute of Technology, Haifa, 3200003, Israel
}}

\title{
{\Large{\textbf{
The Picard-Fuchs equation in classical and quantum physics: Application to higher-order WKB method}}}
}
\date{}

\usepackage[autostyle]{csquotes}
\usepackage[
    backend=biber, 
    sorting=none, 
    style=numeric-comp, 
    url=false
]{biblatex}
\DeclareFieldFormat[article]{volume}{\mkbibbold{#1}} 
\DeclareFieldFormat[article]{number}{(#1)} 
\DeclareFieldFormat{url}{\newline #1} 
\DeclareFieldFormat{doi}{\newline #1} 

\begin{document}
\maketitle
\begin{abstract}

The Picard-Fuchs equation is a powerful mathematical tool which has numerous applications in physics, for it allows to evaluate integrals without resorting to direct integration techniques. We use this equation to calculate both the classical action and the higher-order WKB corrections to it, for the sextic double-well potential and the Lam\'e potential. Our development rests on the fact that the Picard-Fuchs method links an integral to solutions of a differential equation with the energy as a parameter. Employing the same argument we show that each higher-order correction in the WKB series for the quantum action is a combination of the classical action and its derivatives. From this, we obtain a computationally simple method of calculating higher-order quantum-mechanical corrections to the classical action, and demonstrate this by calculating the second-order correction for the sextic and the Lam\'e potential. This paper also serves as a self-consistent guide to the use of the Picard-Fuchs equation.

\end{abstract}
\newpage
\tableofcontents

\section*{Preamble\label{preambuloid}}
\addcontentsline{toc}{section}{\nameref{preambuloid}}

In their seminal work of 1994, Seiberg and Witten
introduced to quantum physics an important concept from the algebraic topology, the Picard-Fuchs equation (PF)~\cite{Seiberg:1994rs}, previously employed by physicists in the context of classical dynamics and integrable systems~\cite{CUSHMAN1985243,Horozov,Dullin1,Dullin2,Dullin3,doi:10.1063/1.4921155,Gulden:2013wya}.
This work marked the beginning of intensive employment of the PF equation in supersymmetric models~\cite{,Ryang:1995hy,Alishahiha:1997bj,Isidro:1996hj,Isidro:1997ut,Isidro:1997jr,Ohta:1999ty,Isidro:2000zw}.

The scope of the current paper consists, at large, of two items. First, we provide a detailed demonstration how to use the PF equation to obtain the classical action for a sextic double-well potential and the Lam\'e potential. With this groundwork accomplished, we present another application of the PF method~--- a simple way to calculate quantum-mechanical corrections to the action.

In mathematics, the PF equation is well-explored in the context of algebraic topology, where it is known as a special case of the Gauss-Manin connection~\cite{EncyMath}. It is a differential equation for periods on a complex manifold, i.e., for integrals along cycles. The principal strength of this entire technique is that it is independent of the coordinates on the manifold, and therefore bears no dependence on a particular path of integration. All such periods must be solutions of the PF equation. Hence, deriving and solving the PF equation allows one to calculate the periods without considering actual integration paths and without performing the brute-force integration. In Appendix~\ref{app:topology}, we discuss this link in more detail; an introduction to the relevant mathematical concepts is also provided in Chapter 2 of~\cite{TGPHD}.

In our paper, we study different applications of the PF equation to two potentials, the sextic double-well potential and the Lam\'e potential. We choose these potentials as instructive examples, because they represent the two main types of spectra in quantum mechanics~--- bound states in an unbounded potential and a band/gap structure in a continuous periodic potential. Both of these potentials are important in the studies of quasi-exactly solvable models, mainly due to the observation of a particular energy spectrum reflection symmetry~\mbox{\cite{Dual,Shifman,ShifmanTurbiner,LONG}}. We apply the PF equation in two different ways. Our approach can be applied to a large variety of problems, therefore we present our calculations in a pedagogical way which allows straightforward application to other potentials. In Section~\ref{sec:firstorder}, we show in great detail how this equation can be used to calculate the classical action, which (to the best of our knowledge) so far has not been written down for these two particular potentials. Performing analytic continuation to complex coordinates, we then write the action as a closed-cycle integral on a complex manifold which is the Riemann surface of the classical momentum. Despite the difference between these potentials' shapes, their Riemann surfaces are topologically equivalent. The classical action is one of the periods on this manifold, for which reason it can be calculated from the solutions of the Picard-Fuchs equation.

In Section~\ref{sec:secondorder}, we extend these ideas and present a technique for calculating the second- and higher-order quantum corrections to the energy levels. These are obtained from the WKB series of the generalised Bohr-Sommerfeld quantisation condition (see Appendix~\ref{app:generalizedBS} for details). Our key observation is that the corrections can be expressed as integrals over a $1$\=/form which is defined on the same manifold as the classical action $1$\=/form $p(x)\d x$. The same tools that were used for deriving the PF equation are now utilized to show that each correction can be written as a combination of derivatives of the classical action. So the computational effort to obtain the quantum corrections to the classical action is much less than the calculation of the action itself. Indeed, the calculation of the corrections is equivalent to deriving the PF equation, but does not require the more involved step of solving the differential equation. This method may also be applied if the classical action is obtained by other means. We demonstrate this by explicitly calculating the second-order corrections; higher corrections at any order can be calculated in a similar way. In Section~\ref{sec:summary} we summarize our results.

When preparing our manuscript for publishing, we learned that similar ideas had been presented in~\cite{Basar:2017hpr} in 2017. In that work, the authors derive a general expression for the higher-order WKB terms for a certain class of genus-1 systems. Their starting point is a quantum generalisation of the classical Wronskian identity for \mbox{genus-$1$} systems. This generalisation implies a relation between the perturbative and non-perturbative terms at all orders in $\hbar$. The authors argue that the quantum corrections for the action in a general genus-1 system can be expressed through the classical action and its first two derivatives. Higher-order derivatives that may appear in the process can be eliminated by using the Picard-Fuchs equation. We approach the problem from a different angle: In the beginning of Section~\ref{sec:secondorder}, we use a geometric argument to show that all the quantum corrections can be expressed through a fixed number of derivatives of the classical action. This number depends only on the genus of the Riemann surface. Starting from this fact, we present a method to calculate quantum corrections at all orders for any potential. While Ba\c{s}ar et al in~\cite{Basar:2017hpr} derive these corrections for a special class of \mbox{genus-$1$} systems, we demonstrate an approach applicable to
all potentials, including those with higher-genus Riemann surfaces. The two examples in this paper are defined on \mbox{genus-$2$} Riemann surfaces. In 2016, the central idea presented in this paper was first reported by one of us in Chapter~5 of~\cite{TGPHD}.

\section{Classical action and first-order WKB}\label{sec:firstorder}

In this section, we calculate the action for two classical potentials by using the Picard-Fuchs equation. The action is defined as
\begin{equation}
    S(E) = \oint \limits_{\mathcal{C}_R} P(x, E)  \d x \quad,
    \label{eq:S}
\end{equation}
an integral over a closed contour (cycle) $\mathcal{C}_R$ in the complex $x$-plane of the momentum\footnote{~The so-defined action is but a Legendre transform of the standard action ${\displaystyle\int\limits^t L(q, \dot{q})\s \d t\s }$, with $L$ being the Lagrangian.}
\begin{equation}\label{eq:p}
    P(x, E) = \sqrt{2\s m\s (E - V(x))}.
\end{equation}
\begin{sloppypar*}
The contour $\mathcal{C}_R$ encloses two real turning points $x_1$ and $x_2$ between which ${V(x) < E}$ and should be close enough to the real axis so that it does not enclose other singularities or branch points. The integral is nonzero because the turning points are the branch points which are connected by a branch cut.\footnote{~Generally speaking, one can freely define the branch cuts. For the real turning points it is usually convenient to make the cut go along the finite interval of the real axis between the turning points.} The momentum is set positive along the bottom of the cut, in order for the action to be positive when the direction of the contour in~\eqref{eq:S} is chosen counter-clockwise. Everywhere else the definition of the momentum follows from analytic continuation.

One application of the classical action is the semiclassical Bohr-Sommerfeld quantisation,
\end{sloppypar*}
\begin{equation}
    S(E_n) = 2\s\pi\s\hbar \left(n+\dfrac{1}{2}\right)\quad.
    \label{eq:BS}
\end{equation}
This condition determines, up to the first order in $\hbar$, the quantum-mechanical energy levels.

\begin{figure}[ht]
\centering
\includegraphics[width=.5\textwidth]{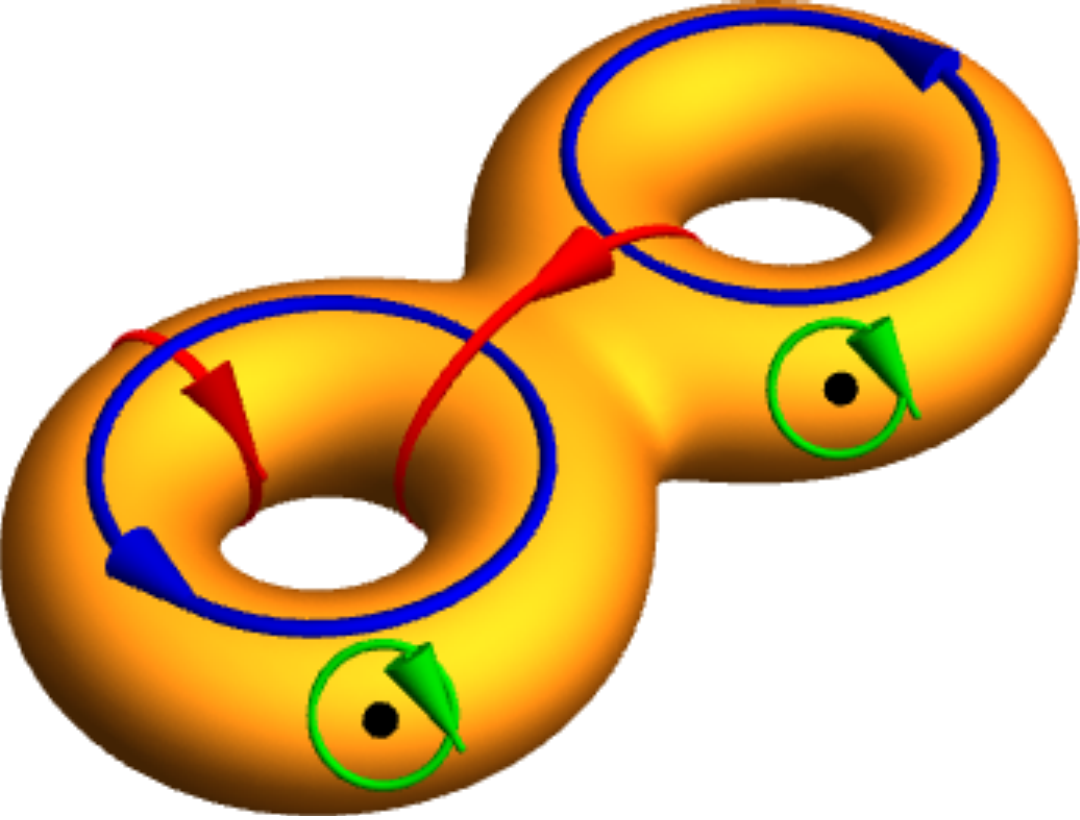}
\caption{The Riemann surface of genus $g=2$ is topologically equivalent to a double torus. There are $2g=4$ linearly independent cycles of integration on this surface~---~one around each handle (blue) and one around each hole (red)~---~plus an additional cycle around the singularity at infinity (green). Any other cycle can be written as a sum of these basis cycles.}
 \label{fig:doubletorus}
\end{figure}

The Picard-Fuchs method is based on the topological properties of the Riemann surface defined by the classical momentum $P(x, E)$ in equation~\eqref{eq:p}.\footnote{~In this work, we only sketch the main steps. For details, we refer the reader to the mathematical literature, e.g.~\cite{EncyMath}. A review, as well as derivations for particular cases of complex manifolds can be found in Chapter 2 of~\cite{TGPHD}.} It is a globally double-valued function of $x$, hence the Riemann surface is constructed out of two copies of the complex plane. These two sheets differ by the choice of sign in front of the square-root. One can travel from one sheet to the other through the branch cuts. This way one obtains a complex manifold of finite genus~$g$ on which one can define a total of~$2g$ linearly independent integration cycles, see Figure~\ref{fig:doubletorus} for an example with~$g=2$. Any other cycle can be deformed into a combination of these $2g$ basis cycles. However, the momentum $P(x,E)$ may additionally have poles resulting in punctures on the manifold, one puncture per sheet for every singularity. Cycles around them are non-trivial, however the cycle around one puncture can be deformed into a sum of the other basis cycles and the cycles around the other punctures. Hence, for $s$ singularities, there are $2s-1$ additional cycles that we need to add to the basis ones.

Thus we have overall $2g+2s-1$ (or $2g$ in the absence of singularities) basis cycles out of which any cycle on the manifold can be constructed. For each cycle $\mathcal{C}_j\,$ we define the \textit{period}~$S_j(E)$ as
\begin{equation}\label{eq:periods}
     S_j(E) = \oint \limits_{\mathcal{C}_j} P(x,E)\d x \equiv \oint \limits_{\mathcal{C}_j} \Lambda(E) \quad.
\end{equation}
Since our main object of interest is the classical action~\eqref{eq:S} it is convenient to choose $\mathcal{C}_R$ as one of the basis cycles.

Our next step is to employ a result from algebraic topology, that on $1$-dimensional complex manifolds (with a finite number of punctures) the number of linearly independent integration cycles is equal to the number of linearly independent $1$\=/forms.\footnote{~Recall that the linear independence of cycles is defined modulo boundary, while the linear dependence of $1$\=/forms is defined up to an exact form. An exact form is a derivative of an analytic function on the manifold, $\d f=\partial_xf(x)\d x$. Integration of this form along any cycle gives zero, ${\displaystyle \oint \limits_C\d f=0}$.} Most importantly, since the number of cycles is finite, the number of basis $1$\=/forms is finite as well, and any other $1$\=/form can be expressed as a linear combination of these. If we take the $1$\=/form ${\Lambda(E)=P(x, E)\d x}$ and start taking derivatives with respect to $E$ we obtain new $1$\=/forms defined on the same manifold. After a finite number $K\leq2g+2s-1$ of steps, we arrive at a $1$\=/form $\Lambda_K(E)=\partial_E^K\Lambda(E)$ which can be expressed as a linear combination of the other derivatives. Or, in other words,
\begin{equation}\label{eq:linearcomb}
 \sum_{k=0}^K \alpha_k\Lambda_k(E) = \d f \quad.
\end{equation}
Upon integration along the closed contour $\mathcal{C}_j$ we obtain
\begin{equation}
 \oint \limits_{\mathcal{C}_j} \Lambda_k(E) =\partial_E^{\s k}S_j(E) \quad,
\end{equation}
while the exact form integrates to zero. This way, condition~\eqref{eq:linearcomb} turns into a differential equation for the periods:
\begin{equation}\label{eq:ODE}
 \sum_{k=0}^K \alpha_k\s\partial_E^{\s k}S_j(E) = 0 \quad.
\end{equation}
This differential equation is known as the Picard-Fuchs equation. The classical action $S(E)$ is a solution to this equation. With the use of physical boundary conditions, we can obtain the classical action from the solutions to this equation. Below we demonstrate how to derive the Picard-Fuchs equation and calculate the action in two cases, the sextic double-well and the periodic Lam\'e potential. While the analytic properties of these potentials are largely different, the corresponding Riemann surfaces are homeomorphic; so the same calculation methods apply.

\subsection{Action of the sextic potential}\label{sec:sextic}

\begin{figure}[ht]
 \centering
 \includegraphics[width=.5\textwidth]{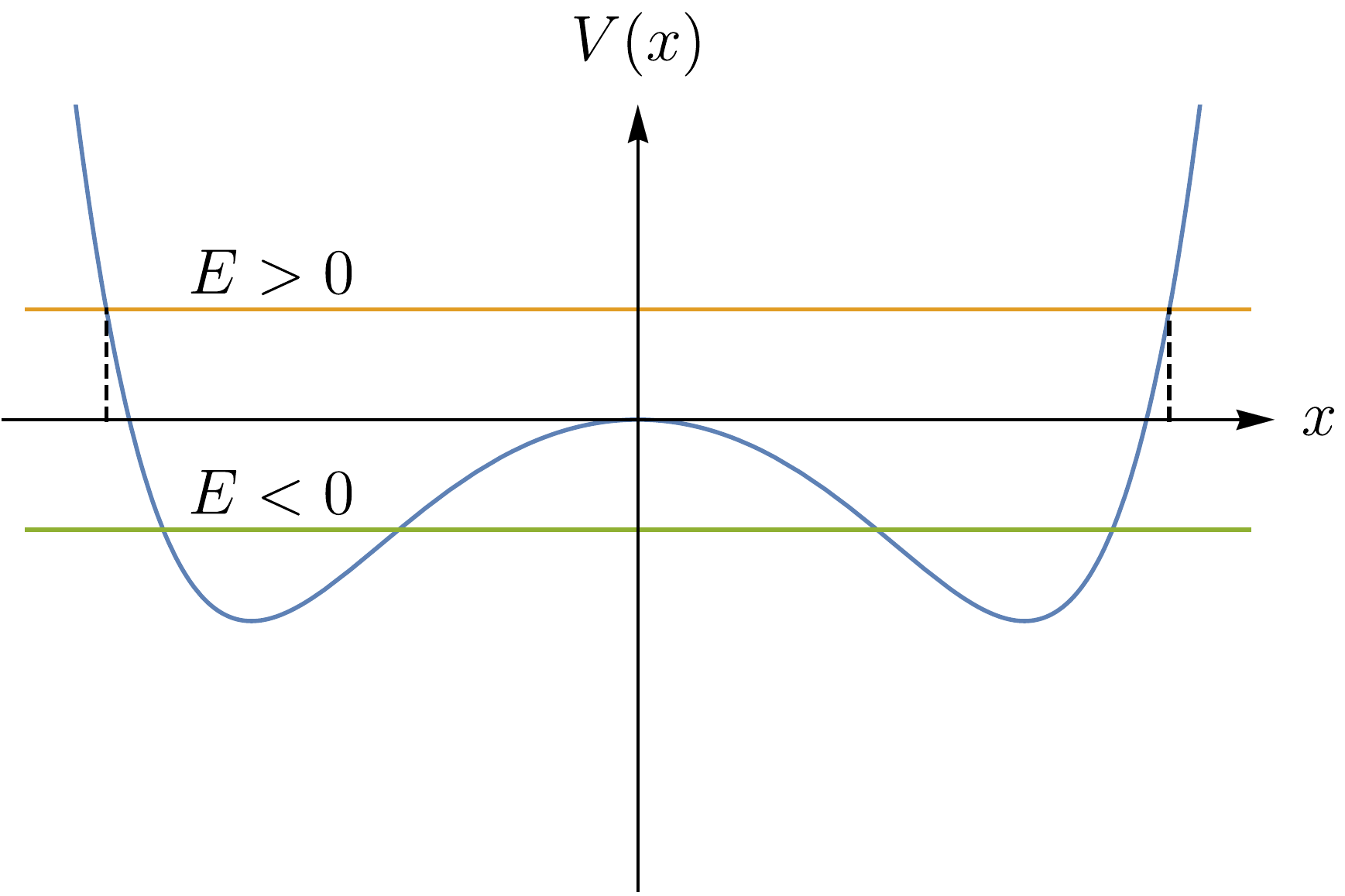}
 \caption{The sextic double-well potential $V(x)$.}
 \label{fig:sexticpot}
\end{figure}

\begin{figure}[ht]
 \centering
 \includegraphics[width=0.45\textwidth]{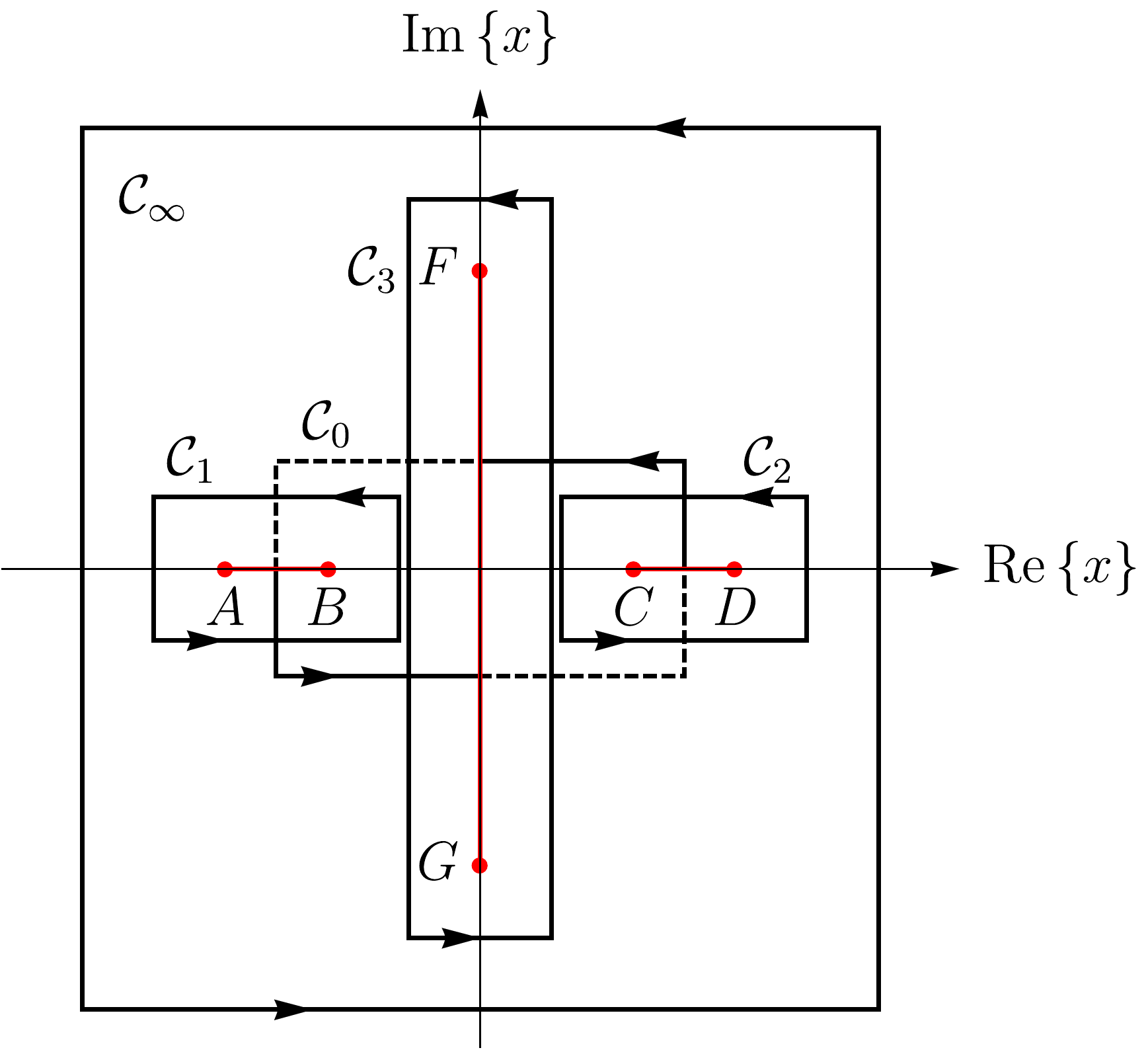}
 \caption{The integration cycles for ${V_{\text{min}} < E < 0}\s$. Solid lines are parts of the cycles that lie on the primary sheet of the Riemann surface, while dashed lines are parts that lie on the second sheet. The primary sheet is identified by taking positive sign in front of the square-root just below the branch cut between $A$ and $B$. The definition everywhere else follows from analytic continuation. Note that while these are the cycles relevant for our calculation, they do not represent a basis, because upon deformation we have: $\mathcal{C}_\infty=\mathcal{C}_1+\mathcal{C}_2+\mathcal{C}_3$.}\label{fig:roots6}
\end{figure}

The first potential we consider is a sextic double-well,
\begin{equation}\label{eq:sexticpot}
    V(x) = - b\s x^2 + d \s x^6 \quad,
\end{equation}
with $b,d\in\mathbb{R}^{\s+}$, see Figure~\ref{fig:sexticpot}. The potential of this shape shows up in the studies of quasi-exactly solvable quantum-mechanical systems, especially because of an observed reflection symmetry of the energy spectrum~\cite{Shifman,ShifmanTurbiner,LONG}. In the following we set the mass to $m=1$ and focus on the energy region $V_{\text{min}}< E < 0$ with
\begin{equation}
    V_{\text{min}}=-\dfrac{2}{3}\sqrt{\dfrac{b^{\s 3}}{3\s d}} \quad.
\end{equation}
For $E>0$ we present the analogous calculation in Appendix~\ref{app:above}. The relevant cycles of integration are shown in Figure~\ref{fig:roots6}, with the corresponding periods defined as in equation~\eqref{eq:periods}.

Upon rescaling the coordinate as $y=(d/b)^{1/4}x$ and the energy as $u=E/V_{\text{min}}$, the momentum~\eqref{eq:p} and the abbreviated action~\eqref{eq:S} take the form
\begin{gather}
    \label{eq:psextic}
    p(y,u) = \sqrt{\dfrac{2}{3^{3/2}}u+y^2-y^6} \qquad,\quad
    s_j(u) =
    \oint \limits_{\mathcal{C}_j} p(y, u) \d y \quad,\\
    S_j(E)
    = \dfrac{\sqrt{2}b^{3/4}}{d^{\s 1/4}}\oint \limits_{\mathcal{C}_j}\sqrt{\dfrac{2}{3^{3/2}}u+y^2-y^6}\d y = \dfrac{\sqrt{2}b^{3/4}}{d^{\s 1/4}}\s  s_j(u)\quad.
    \label{eq:Ssextic}
\end{gather}
\begin{sloppypar*}
Here $s(u)$ is the action of a sextic double-well potential with the two minima at $u=-1$ and the central maximum at $u=0$. The argument of the square-root in~\eqref{eq:psextic} has $6$ roots, therefore there are $6$ branch points which are connected into $3$ branch cuts. This means that the Riemann surface of the momentum $p(y,u)$ has genus $g=2$. These are depicted in Figure~\ref{fig:roots6}, together with the cycles of integration that are important for our analysis. Additionally, $p(y,u)$ has a pole at $y=\infty\,$ which means there is a singularity on each of the sheets, cf. Figure~\ref{fig:doubletorus}. In accordance with the analysis in Section~\ref{sec:firstorder}, there are~$5$ basis cycles and~$5$ linearly independent $1$\=/forms. Hence the Picard-Fuchs equation~\eqref{eq:ODE} is at most of degree~$5$. However, the process of taking derivatives may even sooner yield a $1$\=/form that is linearly dependent of the other derivatives, since there is no guarantee that one can obtain all the basis $1$\=/forms by differentiating ${\lambda(u)=p(y,u)\d y}$.
\end{sloppypar*}

In the case under consideration, the $1$\=/form $p(y,u)\d y$ is even with respect to $y$, i.e. it is symmetric under the change $y \to -y$. However, on the Riemann surface on which these forms are defined, there necessarily exists at least one $1$\=/form that is antisymmetric in $y$. In the de Rham basis,\s\footnote{~For a set of cycles $\mathcal{C}_j\,$, the de Rham basis are the $1$\=/forms $\omega_i$ satisfying ${\displaystyle \oint \limits_{\mathcal{C}_j}\omega_i=\delta_{i,j}}$.} this form will be an antisymmetrised combination of the two $1$\=/forms dual to the cycles $\mathcal{C}_1$ and $\mathcal{C}_2$. Thence the subspace of $1$\=/forms spanned by the derivatives of $\lambda(u)$ is at most $4$-dimensional. Furthermore the residue at the singularity at $y=\infty$ is energy-independent, and we know that ${\displaystyle s_\infty(u)=\oint \limits_{\mathcal{C}_\infty}\lambda(u)=\text{const}}$ also must be a solution of the differential equation~\eqref{eq:ODE}. However, to admit a constant solution, a linear differential equation for~$s(u)$ can't contain a term with~$s(u)$ itself, only the derivatives of $s(u)$. So we look for a linear combination of the form\s\footnote{~Had we not excluded the antisymmetric $1$\=/form from the consideration, we would have ended up with an either undetermined or overdetermined system of equations for coefficients $\alpha_k$ in~\eqref{eq:lincombsextic}.}
\begin{equation}
    \label{eq:lincombsextic}
    \alpha_1\lambda_1 (u) + \alpha_2\lambda_2(u) + \alpha_3\lambda_3(u) + \alpha_4\lambda_4(u) = \d f \quad,
\end{equation}
where $\lambda_k(u)=\partial_u^{\s k}\lambda(u)$. Integration over a closed contour gives a differential equation for the action (cf. equation~\ref{eq:ODE}):
\begin{equation}
    \label{eq:ODEsextic}
    \alpha_1s^{(1)} (u) + \alpha_2s^{(2)}(u) + \alpha_3s^{(3)}(u) + \alpha_4 s^{(4)}(u) = 0 \quad.
\end{equation}
Next we need to find the coefficients $\alpha_n$. It is easy to check that multiplying the left-hand side of~\eqref{eq:lincombsextic} by~$p(y,u)^7$ turns it into a polynomial of order nineteen in $y$. This suggests writing the total derivative on the RHS as:
\begin{equation}
    \label{eq:exactsextic}
    \d f = \partial_y\left[ \dfrac{R_{13}(y)}{p(y,u)^5}\right] \d x
    \quad,\qquad
    \text{where}
    \qquad
    R_{13}(y) = \sum\limits_{k=0}^{13} a_k y^k
    \quad.
\end{equation}
Multiplied by $p(y,u)^7$, the derivative $\d f$ also becomes a nineteenth-order polynomial. Substituting~\eqref{eq:exactsextic} into~\eqref{eq:lincombsextic} and equating coefficients next to the powers of $\s x\s$, we find $a_k$ (which are of no further need) and the desired constants in~\eqref{eq:lincombsextic} (after arbitrarily fixing the overall factor):
\begin{equation}\begin{alignedat}{9}
    \alpha_1 = 5
    \quad,\qquad
    \alpha_2 = 59 u
    \quad,\qquad
    \alpha_3 = 18 (3u^2-1)
    \quad,\qquad
    \alpha_4 = 9 u(u^2-1)
    \quad.
\end{alignedat}\end{equation}
Thus equation~\eqref{eq:ODEsextic} turns into an ordinary differential equation, the Picard-Fuchs equation for the classical action:
\begin{equation}\label{eq:PFsextic}
    5 \s s^{(1)}(u) + 59 \s u\s  s^{\s(2)}(u) + 18\s ( 3 u^2-1)  s^{(3)}(u) + 9\s u\s (u^2 - 1)  s^{(4)}(u) = 0\quad.
\end{equation}
This differential equation is of the generalized hypergeometric type~\cite{NIST16}, so its basis solutions can be expressed through the generalized hypergeometric functions $_pF_q\,$. Using software for solving differential equations we find:
\begin{equation}\begin{alignedat}{9}\label{eq:Solsextic}
 F_0(u) &=  1\quad&&,\\
 F_1(u) &= u \s _3F_2\left(\{\frac{1}{6},\frac{1}{2},\frac{5}{6}\},\{1,\frac{2}{3}\};u^2\right)\quad&&,\\
 F_2(u) &= u^2 \s _4F_3\left(\{\frac{2}{3},1,1,\frac{4}{3}\},\{\frac{3}{2},\frac{3}{2},2\};u^2\right)\quad&&,\\
 F_3(u) &= \Gamma\left(\frac{1}{6}\right)^3u^{2/3} \s _3F_2\left(\{\frac{-1}{3},\frac{1}{6},\frac{1}{6}\},\{\frac{1}{3},\frac{2}{3}\};\frac{1}{u^2}\right)\quad&&,\\
         &+ 2^{1/3}3\pi^{3/2}u^{-2/3} \s _3F_2\left(\{\frac{1}{3},\frac{5}{6},\frac{5}{6}\},\{\frac{4}{3},\frac{5}{3}\};\frac{1}{u^2}\right)\quad&&,
\end{alignedat}\end{equation}
where $\Gamma$ is the factorial gamma function.\footnote{~The definitions of the $_pF_q$ and $\Gamma$ functions are taken as in $\,$\url{mathworld.wolfram.com}~\cite{mathworld}.}
Every period in equation~\eqref{eq:periods} is an integral on the manifold, so it must be a linear combination of these basis solutions,
\begin{equation}
    \label{eq:actionsextic}
     s_j(u) = \sum_{k=0}^3C_{j,k}F_k(u) \quad.
\end{equation}
The cycles which are relevant for our calculations are shown in Figure~\ref{fig:roots6}, where we know from symmetry $ s_1(u)= s_2(u)$. In order to identify the coefficients $C_{j,k}\,$ we use analytic properties of the basis solutions~\eqref{eq:Solsextic} near critical values of the parameter $u$, for which one or more of the integration cycles shrink to a point.

We begin with calculating the action for the case of $-1<u<0$, the case of $u>0$ is shown in Appendix~\ref{app:above}. To this end we first consider the auxiliary cycle $\mathcal{C}_0$ which encloses the two branch points B and C in Figure~\ref{fig:roots6}.  As $u\to0\s$, this cycle shrinks to a point. Therefore any integral along this cycle goes to zero and is analytic in $u$. $F_3(u)$ is non-analytic at $u=0$, thus it can't be a part of $ s_0(u)$ and $C_{0,3}=0$. Similarly, because $F_0(u)$ does not vanish we obtain $C_{0,0}=0$.

\begin{sloppypar*}
To find the other two coefficients we note that ${F_1(u)=u+\mathcal{O}(u^3)}$ and ${F_2(u)=u^2+\mathcal{O}(u^4)}$. Expanding the integrand $\lambda(u)$ in powers of $u$ and performing a residue calculation for each term we arrive at
\end{sloppypar*}
\begin{equation}
  s_0(u) = \dfrac{2\pi \iu}{3^{3/2}}u + \mathcal{O}(u^3) \quad.
\end{equation}
\begin{sloppypar*}
There is no quadratic term here, which means that $C_{0,2}=0$, and the linear term gives us ${C_{0,1}=\dfrac{2\pi \iu}{3^{3/2}}}$. This period is therefore
\end{sloppypar*}
\begin{equation}
     s_0(u) = \dfrac{2\pi \iu}{3^{3/2}}F_1(u)\quad.
    \label{eq:sexticS0}
\end{equation}
From the quantum-mechanical point of view this is the tunneling action which defines the non-perturbative corrections to the energy levels within the semi-classical approximation~\cite{LL}. We do not dwell on these corrections here, but will need this period below.

\begin{figure}[ht]
 \includegraphics[width=\textwidth]{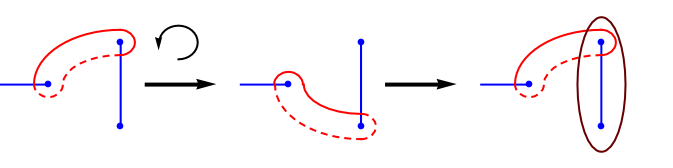}
 \caption{Exchange of two branch points (blue) under the monodromy $u\to u \s \me^{2\pi \iu}$. The branch points move, but the integration cycle (red) may never cross a branch point and therefore is dragged along. To restore the original cycle, one needs to add an additional cycle around the two branch points which were exchanged (maroon).}
 \label{fig:monodromy}
\end{figure}

\begin{sloppypar*}
Now we have everything to calculate the classical action $ s_1(u)$. There are two special values for the parameter: $u=-1$ (this is where the cycle $\mathcal{C}_1$ contracts to a point) and $u=0$ (this is where the branch points $B$ and $C$ merge). In the latter case the action $ s_1(u)$ is non-analytic. To see this we employ a concept from algebraic topology called {\em monodromy}. We perform a rotation of the parameter $u$ in the complex plane around the critical value, i.e., $u\to u \s\me^{2\pi \iu}$. In the end the structure of branch points is the same as before. However, in the process the points $B$ and $C$ swap positions, see Figure~\ref{fig:monodromy}. Such a monodromy causes a deformation of the cycles, which is equivalent to adding a cycle around the points B and C to $\mathcal{C}_1$~\cite{Monodromy,TGPHD}. In our words this means ${\mathcal{C}_1\to \mathcal{C}_1+\mathcal{C}_0}$. Likewise, the integral along that cycle has to obtain the same additional term, ${ s_1(u\s\me^{2\pi \iu}) =  s_1(u)+ s_0(u)}$. The only function which, upon changing its argument by $\me^{2\pi \iu}$, acquires an additive term is the complex logarithm. Therefore we can write
\end{sloppypar*}
\begin{equation}
     s_1(u) = Q_1(u) + \dfrac{ s_0(u)}{2\pi \iu}\log(u)\quad,
\end{equation}
where $Q_1(u)$ is analytic near $u=0$. Among the solutions \eqref{eq:Solsextic},
the function $F_3(u)$ is the only solution which is non-analytic near the origin. We expand it near the origin up to the $u\log(u)$ term and compare it with $\dfrac{ s_0(u)}{2\pi \iu}\log(u)$ to get
\begin{equation}
   \label{eq:sexticC13}
    C_{1,3}=\left[-12\s\Gamma\left(\dfrac{1}{3}\right)\Gamma\left(\dfrac{1}{6}\right)\right]^{-1}\quad.
\end{equation}
\begin{sloppypar*}
Additionally, at $u=0$ the integral can be evaluated analytically: ${ s_1(0) = \dfrac{\pi}{4}}$. At ${u=0}$, only $F_0(u)$ and $F_3(u)$ are non-zero. Using the value $F_3(0)=\pi\s\Gamma\left(\dfrac{1}{3}\right)\Gamma\left(\dfrac{1}{6}\right)$ and $C_{1,3}$ from~\eqref{eq:sexticC13}, we calculate the coefficient of $F_0(u)$:
\end{sloppypar*}
\begin{equation}
    \label{eq:sexticC10}
    C_{1,0}=\dfrac{\pi}{3}\quad.
\end{equation}
\begin{sloppypar*}
We now turn to ${u=-1}$ where the cycle $\mathcal{C}_1$ contracts to a point. This contraction has two consequences:
\end{sloppypar*}
\begin{enumerate}
    \item $ s_1(u)\to0$ as $u\to-1$,
    \item $ s_1(u)$ is analytic around $u=-1$.
\end{enumerate}
\begin{sloppypar*}
From these two constraints and the coefficients we have already obtained, we can calculate the coefficients $C_{1,1}$ and $C_{1,2}$. We expand all the solutions $F_j(u)$ around $u=-1$, and require that the constant terms and the leading non-analytic terms ${(u+1)\log(u+1)}$ vanish in equation~\eqref{eq:actionsextic} for $s_1(u)$. We find that both conditions are met only if
\end{sloppypar*}
\begin{equation}
   \label{eq:sexticC11C12}
   C_{1,1} = C_{1,2} = 0 \quad.
\end{equation}
The result for the classical action in \eqref{eq:actionsextic} therefore is
\begin{equation}\label{eq:actionsexticbelow}
    s_1(u) = \frac{\pi}{3} - \left[12\s\Gamma\left(\dfrac{1}{3}\right)\Gamma\left(\dfrac{1}{6}\right)\right]^{-1} \!\!F_3(u)\quad.
\end{equation}

\begin{sloppypar*}
In Appendix~\ref{app:above}, we show the explicit calculation of the classical action~$ \tilde{s}_3 (u)$ above the local maximum at ${u=0}$. As an independent check of our calculations, we can verify that the actions obey the duality property which was pointed out in our previous work~\cite{LONG}:
\end{sloppypar*}
\begin{equation}
    2 s_1(-u) + \tilde{s}_3 (u)
    = \pi \quad,\qquad 0<u<1\quad.
    \label{eq:dualsextic}
\end{equation}
It is easy to show that the results in equations~\eqref{eq:actionsexticbelow} and~\eqref{eq:actionsexticabove} satisfy this property.

\begin{figure}[ht]
\centering
\begin{minipage}{1\textwidth}
 \centering
\includegraphics[width=1\textwidth]{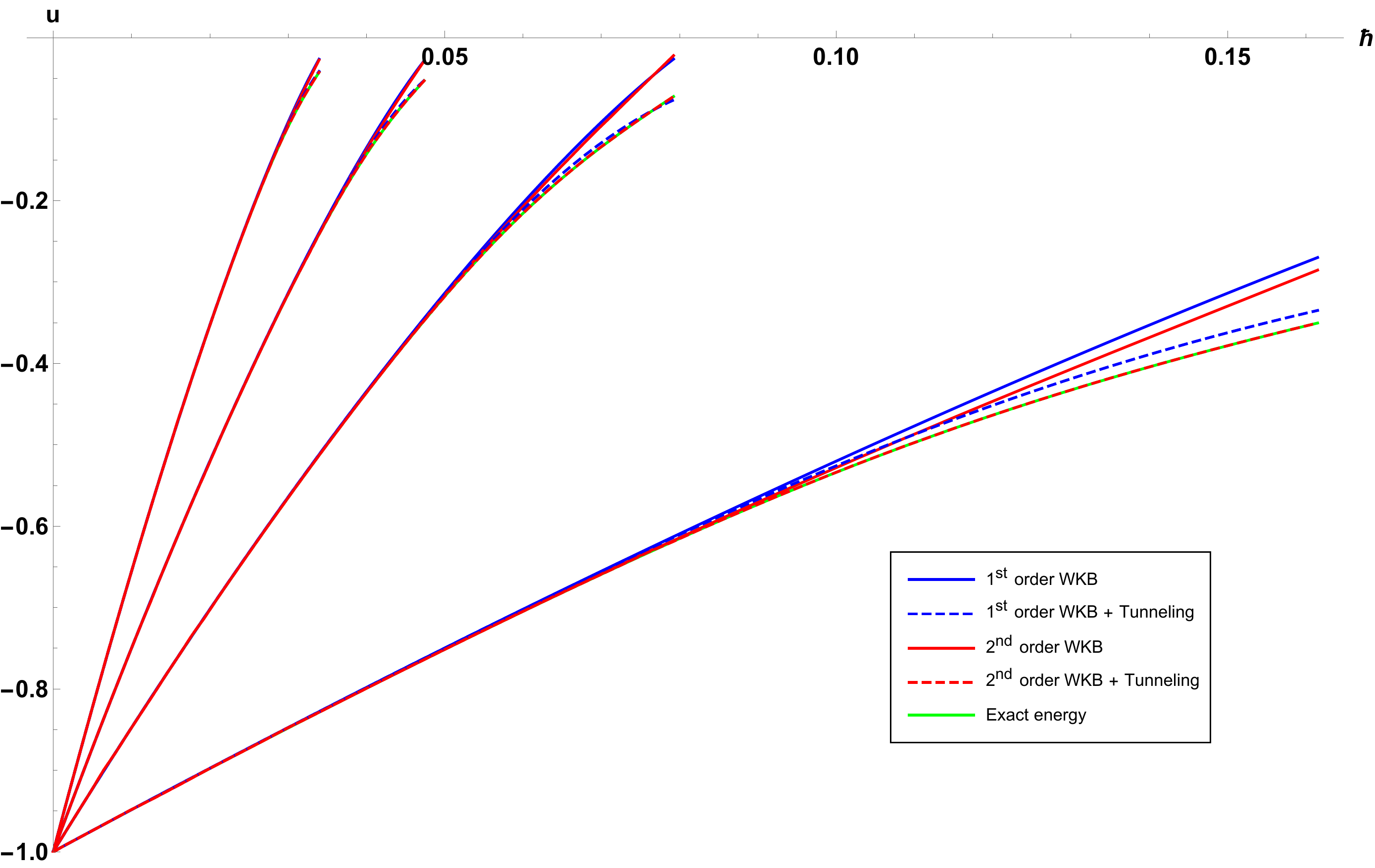}
\end{minipage}
\caption{Energy states in the double-well potential: the four lowest levels below the local maximum at $u=0$. The results obtained from the first-order WKB are shown in blue, second-order in red; an exact numerical result is given in green. The dashed lines show the results obtained with the tunneling corrections taken into account. Below the maximum, the eigenstates appear in pairs which are split by tunneling effects. For the numerical result, we only show the lower eigenstate.}\label{fig:sexticspec}
\end{figure}

The afore-derived actions can be employed to find the spectrum of the corresponding quantum-mechanical system. For the levels below and above the maximum, the Bohr-Sommerfeld quantisation condition reads:
\begin{subequations}\label{bz}
\begin{alignat}{9}
     s_1(u_n) &= 2 \s \pi \s \hbar \s \left( n + \dfrac{1}{2} \right) \quad&&, \qquad -&&1&&<u&&<0 \quad&&,\\
     \tilde{s}_3(u_n) &= 2 \s \pi \s \hbar \s \left( n + \dfrac{1}{2} \right) \quad&&, \qquad &&0&&<u&&<1 \quad&&.
\end{alignat}
\end{subequations}

To find $\s u_n\s$, we solve those equations numerically. Figure~\ref{fig:sexticspec} shows the first four energy levels as functions of $\hbar$, below and above the local maximum. The actual values of the energy levels are well approximated by their Bohr-Sommerfeld counterparts. The latter are virtually indistinguishable from the results rendered by second-order WKB calculation with the tunneling effects taken into account. To the one-instanton order, these effects are described by the action $s_0(u)$, see~\cite{Garg} for a detailed discussion.

\subsection{Action of the elliptic potential}\label{sec:Lame}

\begin{figure}[ht]
\centering
\begin{minipage}{0.45\textwidth}
\centering
\includegraphics[width=1\textwidth]{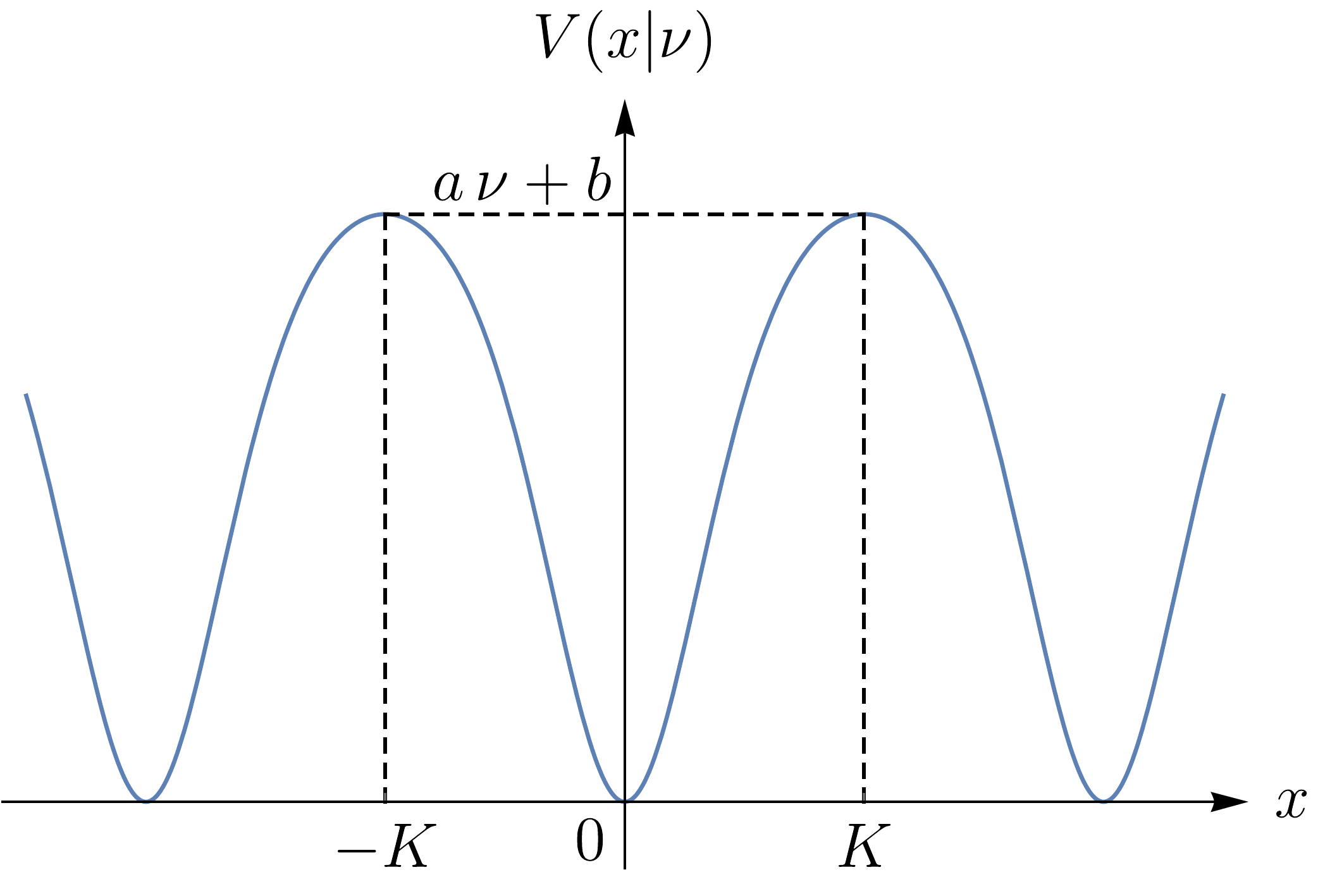}
\caption{The periodic potential $V(x|\nu)$.}\label{fig:Lame}
\end{minipage}\hfill
\begin{minipage}{0.45\textwidth}
\centering
\includegraphics[width=1\textwidth]{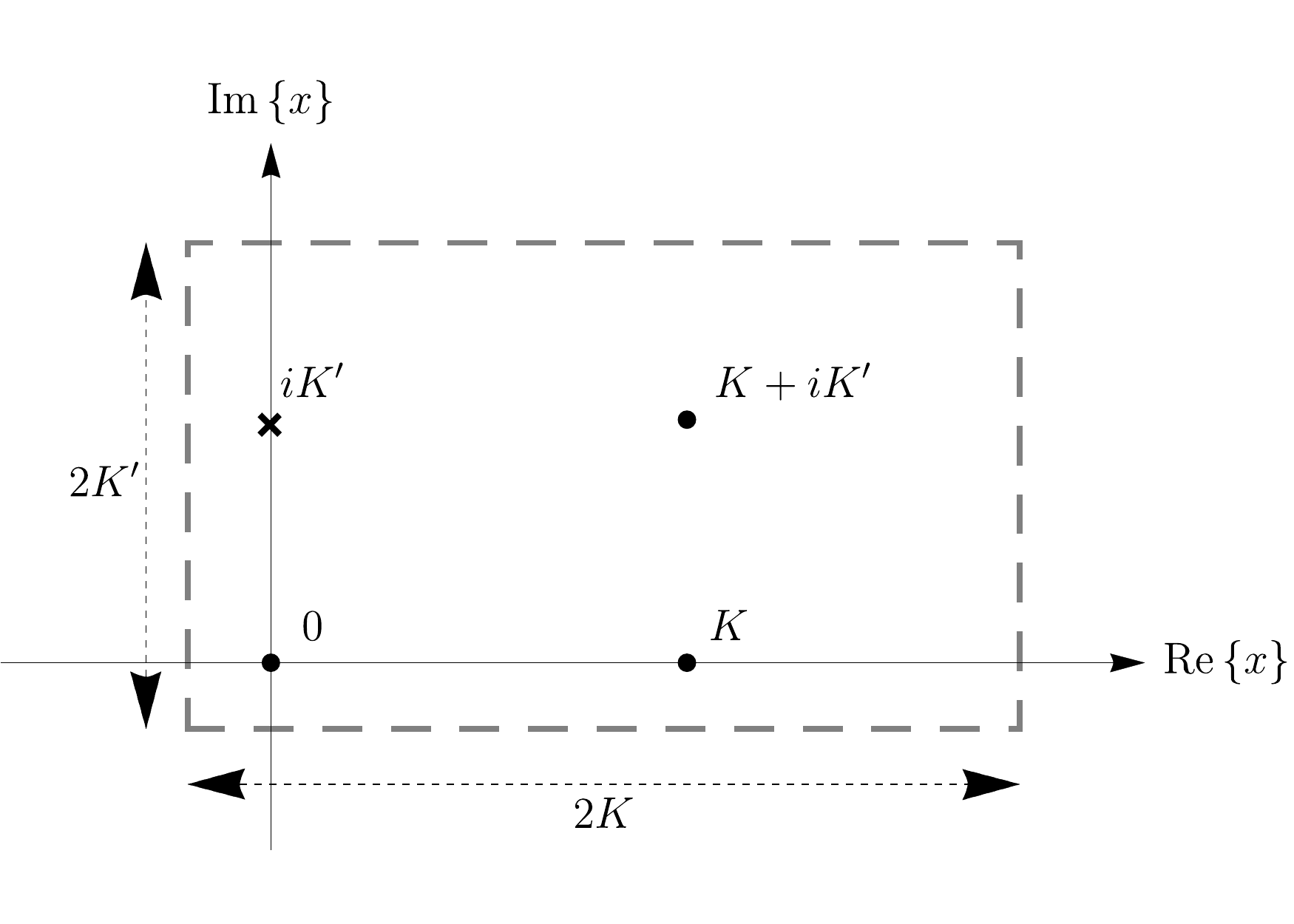}
\caption{The fundamental parallelogram of the potential $V(x|\nu)$.}\label{fig:parallelogram}
\end{minipage}
\end{figure}

As a second example, we calculate the action for a periodic potential defined in terms of the Jacobi elliptic function $\sn(x,\nu)$~\cite{NIST}:
\begin{equation}\begin{gathered}
 V(x|\nu) = a\s\nu \s\snt (x| \nu) + b \quad,\\
 a \in \mathbb{R}^{\s+}\quad,\qquad b \in \mathbb{R} \quad,\qquad\nu \in [0,1]\quad,
 \label{eq:Ve}
\end{gathered}\end{equation}
\begin{sloppypar*}
For a special choice of the constant~$a$, potential~\eqref{eq:Ve} turns into the widely studied Lam\'e potential~\cite{ars,akhi,susy,Dual}. It is a doubly periodic meromorphic function whose real and imaginary periods are $2K(\nu)$ and ${2\iu K'(\nu) \equiv 2\iu K(1-\nu)}$, with
\end{sloppypar*}
\begin{equation}
    K(\nu) = \int \limits_0^{\pi/2}
    \dfrac{\d\theta}{\sqrt{1-\nu\s \sin^2 \theta}}
\end{equation}
being the complete elliptic integral of the first kind. The potential and its fundamental parallelogram are shown in Figures~\ref{fig:Lame} and~\ref{fig:parallelogram}. %

\begin{figure}[ht]
 \centering
 \includegraphics[width=.7\textwidth]{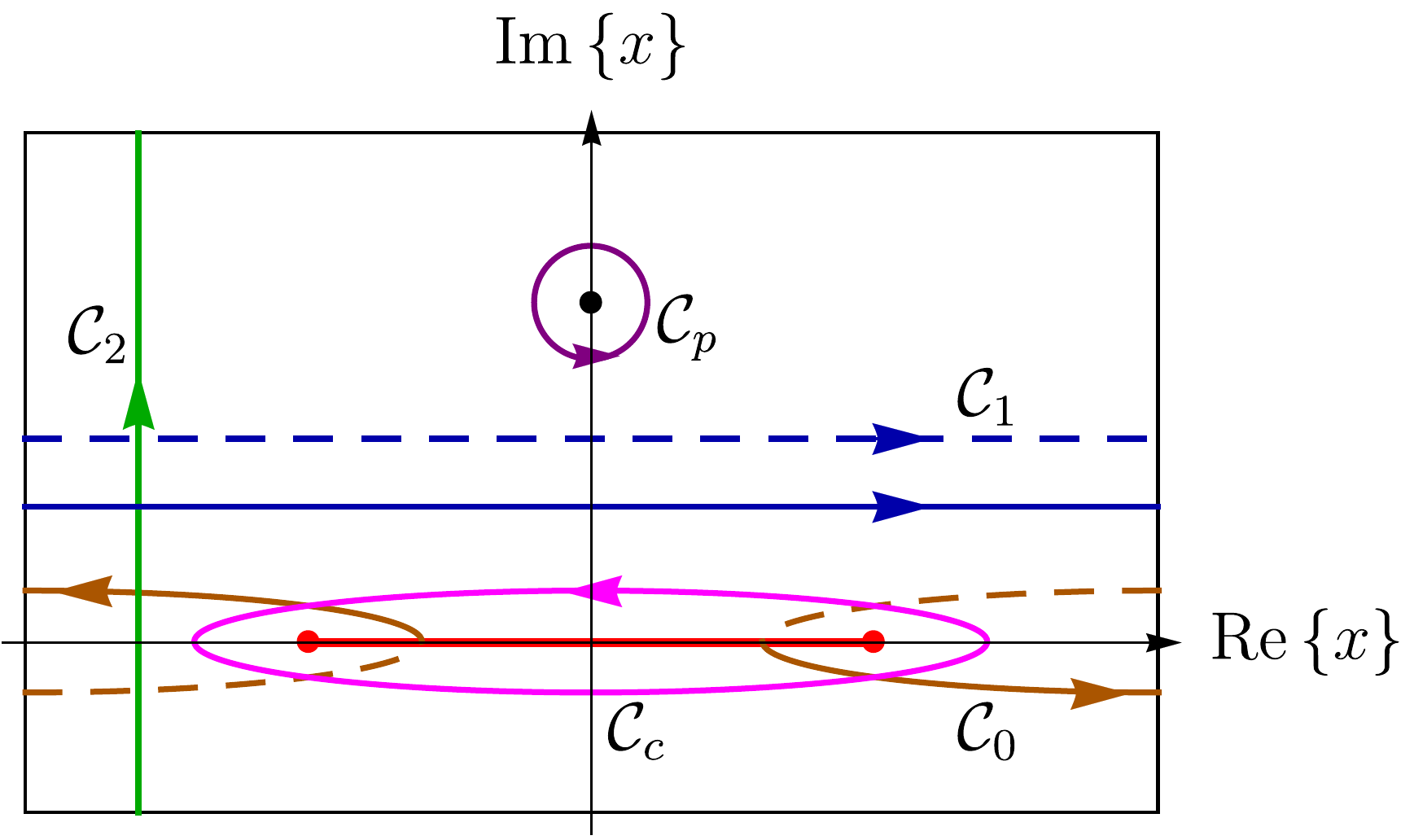}
 \caption{The basis cycles on the Riemann surface of the momentum in equation~\eqref{eq:plame}. The red line is the branch cut between the classical turning points. The period of the cycle $\mathcal{C}_c$ around them  gives the classical action.  The parts of the contours, which are denoted with solid lines, lie on the first sheet. The parts denoted with dashed lines are on the second. Note that the trajectories $\mathcal{C}_1$ and $\mathcal{C}_2$ are closed contours due to the periodicity. The trajectory $\mathcal{C}_1$ (blue) has to travel across both sheets (solid \textit{and} dashed) to be closed.}
 \label{fig:LameRS}
\end{figure}

\begin{sloppypar*}
In the following, we rescale the energy as $\,{u= \dfrac{E-b}{a\s\nu}-1}\,$ and choose the mass as ${\s m=1/2\s}$. Then the canonical momentum and the abbreviated action take form
\end{sloppypar*}
\begin{gather}
    \label{eq:plame}
    p (x,u|\nu) = \sqrt{u + \cnt (x|\nu)}\quad,\qquad
     s_j(u|\nu) = \oint \limits_{\mathcal{C}_j} p (x,u|\nu) \d x \equiv \oint \limits_{\mathcal{C}_j} \mu(u|\nu) \quad,\\
    \label{eq:Slame}
    S_j(E|\nu) = \oint \limits_{\mathcal{C}_j}\sqrt{E-a\nu\snt(x|\nu)-b} \d x = \sqrt{a \nu} \oint\limits_{\mathcal{C}_j} \sqrt{u + \cnt(x,\nu)} \d x= \sqrt{a \nu}\s  s_j(u|\nu)\quad,
\end{gather}
\begin{sloppypar*}
where ${\cnt(x|\nu) = 1-\snt(x|\nu)}$ and the cycles $\mathcal{C}_j$ are defined in Figure~\ref{fig:LameRS}. In the main text we focus on the energy regime between the minimum and maximum of the potential, ${-1<u<0}$. In Appendix~\ref{app:above}, we present the calculation for the classical action above the maximum.
\end{sloppypar*}

\begin{sloppypar*}
The double periodicity of  potential~\eqref{eq:Ve} carries over to  momentum~\eqref{eq:plame}. However, owing to the presence of the square root, the period in the real direction is doubled and equals ${4 K(\nu)}$. Thus the momentum $p(x,u|\nu)$ is a doubly periodic function whose real and imaginary periods are ${4 K(\nu)}$ and ${2\iu K'(\nu)}$, correspondingly. For analytic continuation of the momentum we need to choose the opposite sign in the second copy of the fundamental parallelogram, i.e. this is the second sheet of the Riemann surface. Gluing together the edges of the parallelogram gives the topological structure of a torus. Additionally, the two zeroes of the argument of the square root are branch points, for ${-1<u<0}$ they are the classical turning points. We connect these into a branch cut, crossing it also allows to travel between the two sheets. For the Riemann surface, this means that we have to cut it open along the cuts and connect the edges to the opposite sheet. This transforms the torus into a double torus, akin to the example of the sextic potential. As before the potential has one pole, at ${x=i K'(\nu)}$, which appears on both sheets of the Riemann surface. Hence the Riemann surface is a manifold of genus ${g=2}$ with two singularities, cf. Figure~\ref{fig:doubletorus}. This means that, as in Section~\ref{sec:sextic}, there are $5$ independent integration cycles and $5$ independent $1$\=/forms.\,\footnote{~It is noteworthy that while the two potentials are very different in their physical and analytic properties, the underlying Riemann surfaces share the same topology. This is no surprise since the Riemann surface is constructed of two sheets and is topologically equivalent to a multi-torus with a certain number of singularities.} Figure~\ref{fig:LameRS} shows the fundamental parallelogram with a full set of linearly independent basis cycles.
\end{sloppypar*}

We are now in a position to derive the Picard-Fuchs equation for the elliptic potential. Since the Riemann surface in this case is again of genus~$2$ and has two punctured points, the degree of the Picard-Fuchs equations is, at most,~$5$. However, in reality the degree is, at most,~$3$.\s\footnote{
~Thus, there is no contradiction with the fact that all genus-$1$ surfaces can be described in terms of elliptic functions; see~\cite{ww,bateman}, and also~\cite{Basar:2017hpr}.
}  To demonstrate this, we recall that the $1$\=/form $\mu(u|\nu)$, as well as each of its derivatives with respect to $u$, has opposite signs on the two sheets of the Riemann surface. Therefore, integrating it along a trajectory which has equal portions on both sheets gives zero. Trivially, the cycle $\mathcal{C}_1$ has this property, wherefore its dual $1$\=/form cannot be expressed via $\mu(u|\nu)$ and its derivatives. Furthermore we define the cycle $\mathcal{C}_c'$ as the same as the cycle $\mathcal{C}_c$ but on the second sheet, i.e. with a dashed line in Figure~\ref{fig:LameRS}. Then integrating $\mu(u|\nu)$ or any of its derivatives along the cycle $\tilde{\mathcal{C}}_c + \mathcal{C}_c$ also gives zero. In terms of the basis cycles this can be expressed as
\begin{equation}
 \mathcal{C}_c'+\mathcal{C}_c=2\s\mathcal{C}_c+\mathcal{C}_2+\mathcal{C}_p \quad.
\end{equation}
\begin{sloppypar*}
Hence the $1$\=/form dual to this combination cannot be obtained from $1$\=/forms that are generated from $\mu(u|\nu)$. For this reason there exist at most $3$ linearly independent $1$\=/forms that can be obtained by differentiation of $\mu(u|\nu)$ with respect to $\,u\,$. We denote them as ${\mu_k (u|\nu) \equiv \partial_u^{\s k} \mu_k (u|\nu)}$.
Furthermore, the residue at the pole is independent of $u$. This means that a differential equation for the periods with respect to $u$ must admit a constant solution and therefore cannot contain $s(u)$, only its derivatives. All in all, we see that there exists a linear combination of the first three derivatives which equals an exact form:
\end{sloppypar*}
\begin{equation}
 \label{eq:lincombLame}
    \beta_1\s \mu_1 (u|\nu) + \beta_2\s \mu_2(u|\nu) + \beta_3 \s\mu_3(u|\nu)
    = \d g \quad,
\end{equation}
 \begin{sloppypar*}
integration whereof leads us to the Picard-Fuchs equation:
\end{sloppypar*}
\begin{equation}
    \beta_1  s^{(1)}(u|\nu) + \beta_2  s^{(2)}(u|\nu) + \beta_3  s^{(3)}(u|\nu) = 0 \quad.
\end{equation}
Evaluating the derivatives on the left-hand side of equation~\eqref{eq:lincombLame} and multiplying these by $p(x,u|\nu)^5$, we arrive at a fourth-order polynomial in $\cn(x|\nu)$, with only even powers. To match this, we need to design an exact form with the same property, which we find as
\begin{equation}\label{eq:exactLame}
    \d g = \partial_x\left[\dfrac{\cn(x|\nu)\sn(x|\nu)\dn(x|\nu)}{p(x,u|\nu)^3}\right]\d x \quad.
\end{equation}
Here $\cn(x|\nu)$, $\sn(x|\nu)$, and $\dn(x|\nu)$ are the Jacobi elliptic functions~\cite{mathworld}. The choice in~\eqref{eq:exactLame} is guided by the properties of the elliptic functions:
the product of $\d g$ by $p(x,u|\nu)^5$ contains the even powers of elliptic functions solely. All of those can be expressed via $\cnt(x|\nu)$~\cite{mathworld}.
Then, multiplying equation~\eqref{eq:lincombLame} by $p(x,u|\nu)^5$, we obtain:
\begin{equation}\begin{multlined}
    \dfrac{1}{8} \bigl(3 \beta_3-2 \beta_2 u +4 \beta_1 u^2 +2 (4 \beta_1 u - \beta_2) \cnt(x|\nu )+4\beta_1 \cntt(x|\nu )\bigr) \\
    =1 + u -2 (u +\nu + u\s\nu) \left(1-\cnt (x|\nu)\right) + (\nu(2+3\s u)-1) \left(1-\cnt (x|\nu)\right)^2 \quad.
\end{multlined}\end{equation}
Equating coefficients next to the powers of $\cnt(x|\nu)$, one finds:
\begin{equation}\begin{gathered}
    \beta_1 = 3\nu u+ 2\nu-1
    \quad,\qquad
    \beta_2 = 4(\nu u(3u+4)+\nu-2u-1)
    \quad,\\
    \beta_3 = 4u(1+u) (\nu u+\nu-1)
    \quad.
\end{gathered}\end{equation}
Thus, the Picard-Fuchs equation for the action $ s(u|\nu)$ is:
\begin{equation}\begin{multlined}
    \label{eq:PFLame}
    \left(3\s\nu\s u+ 2\s\nu - 1\right)  s^{(1)}(u|\nu)
    +
    4\left(\nu\s u(3\s u+4)+\nu-2\s u-1\right)  s^{(2)}(u|\nu)
    \\+
    4\s u \left(1+u\right) \left(\nu\s u + \nu - 1\right)   s^{(3)}(u|\nu) = 0\quad.
\end{multlined}\end{equation}
The basis solutions to this equation are
\begin{equation}\label{eq:solLame}
  \begin{alignedat}{9}
        G_0 &= 1 \quad&&,\\
        G_1(u,\s u_0|\nu) &=  \int \limits_{u_0}^u \d v \dfrac{\lP_{-1/2}\left(\dfrac{(v+1)\nu-1-2v}{(v+1)\nu-1}\right)}{\sqrt{(v+1)\nu-1}} \quad&&,\\
        G_2(u,\s u_0|\nu) &=  \int \limits_{u_0}^u \d v \dfrac{\lQ_{-1/2}\left(\dfrac{(v+1)\nu-1-2v}{(v+1)\nu-1}\right)}{\sqrt{(v+1)\nu-1}} \quad&&,
  \end{alignedat}
\end{equation}
with $\lP_n(u)$ and $\lQ_n(u)$ being the Legendre polynomials of order $n$ of the first and second kind, respectively~\cite{mathworld}.

\begin{sloppypar*}
From there, we want to find the classical action below the maximum of the potential, ${-1<u<0}$. In the following it is convenient to choose the integration limit at the minimum of the potential, ${u_0=-1}$. In terms of the basis solutions~\eqref{eq:solLame}, the action assumes the form
\end{sloppypar*}
\begin{equation}
  s_c(u|\nu) = D_{0}G_0 + D_{1}G_1(u,-1|\nu) + D_{2}G_2(u,-1|\nu) \quad.
\end{equation}
To calculate the coefficients $D_k\,$, we need to obtain three conditions on them. To this end, consider the behaviour of the action near the minimum of the potential, where we can show that
\begin{enumerate}
    \item $ s_c(u|\nu)$ is analytic as $u\to-1$\s,
    \item $ s_c(u|\nu) \to 0$ as $u\to-1$\s,
    \item $\left.\partial_u^{\s+} s_c(u|\nu)\right|_{u=-1}=\pi$, \,where $\partial_{u}^{\s+}$ is the right derivative with respect to $u$\s.
\end{enumerate}
\begin{sloppypar*}
The first two conditions stem from the fact that the cycle $\mathcal{C}_c$ contracts to a point as ${u\rightarrow-1}$, a situation similar to the one discussed in Section~\ref{sec:sextic}. The third condition is obtained by the direct evaluation of the derivative of the action integral:
\end{sloppypar*}
\begin{equation}
\begin{multlined}
    \partial_{u}^{\s+} s_c(u|\nu)\Bigl.\Bigr\rvert_{u=-1}
    = \partial_u^{\s+} \oint\limits_{\mathcal{C}_c} \sqrt{u + \cnt(x|\nu)} \d x \Bigl.\Bigr\rvert_{u=-1}
    = \oint\limits_{\mathcal{C}_c} \dfrac{1}{2\sqrt{-1+ \cnt(x|\nu)}} \d x
    \\= 2\pi\iu\Res\left\{\dfrac{1}{2\s\iu\sn(x|\nu)}, x=0\right\} =\pi\quad.
\end{multlined}
\end{equation}
The function $F_1(u,u_0|\nu)$ is non-analytic when $u\to-1$, which implies $D_{1}=0$. To find the two other coefficients, we use the two remaining conditions:
\begin{equation}
    \left\{\begin{alignedat}{9}
    D_0 + D_2 \s G_2(-1,-1|\nu) &= 0 \\ D_2 \, \partial_u^{\s+} G_2 (u,-1|\nu) \Bigl.\Bigr\rvert_{u=-1} &= \pi
    \end{alignedat}\right.
    \qquad \Longrightarrow \qquad
    \left\{\begin{alignedat}{9}
    D_0 &= 0\\D_2 &= 2\iu
    \end{alignedat}\right.
    \quad.
\end{equation}
Hence, we obtain for the classical action of the Lam\'e potential:
\begin{equation}
    \label{eq:actionLame}
     s_c(u|\nu) = 2\iu G_2(u,-1|\nu) = 2\iu \int \limits_{-1}^u \d v \dfrac{\lQ_{-1/2}\left(\dfrac{(v+1)\nu-1-2v}{(v+1)\nu-1}\right)}{\sqrt{(v+1)\nu-1}}
    \quad.
\end{equation}
In Appendix~\ref{app:above}, we perform a similar calculation to obtain the classical action above the maximum of the potential, i.e., for the unbounded motion in the periodic potential:
\begin{equation}
    \label{eq:actionLame2}
    \tilde{s}_c(u|\nu) = \dfrac{\pi}{\sqrt{\nu}} - \dfrac{\pi}{2} G_1 \biggl(\biggl.u,\dfrac{1-\nu}{\nu}\biggr|\nu\biggr) + \iu G_2 \biggl(\biggl.u,\dfrac{1-\nu}{\nu}\biggr|\nu\biggr)
    \quad.
\end{equation}
Using analytic properties of the Legendre polynomials~\cite{mathworld}, one can cast this expression as
\begin{equation}
    \tilde{s}_c(u|\nu) = \dfrac{\pi}{\sqrt{\nu}} - \iu\sqrt{\dfrac{1-\nu}{\nu}} G_2 \biggl(\biggl.-u\dfrac{\nu}{1-\nu},-1\biggr|1-\nu\biggr)
    \quad.
\end{equation}
From this, it is easy to confirm a duality property for the action that was first derived in~\cite{LONG},
\begin{equation}
    2 \sqrt{1-\nu}\tilde{s}_c\biggl(\biggl.u\dfrac{\nu}{1-\nu}\biggr|1-\nu\biggr) + \sqrt{\nu}s_c(-u|\nu) = 2\pi \quad.
\end{equation}
This serves as an additional confirmation of our result.

For completeness, we also show the derivation of the instanton action. It can be obtained by integration over the cycle $\mathcal{C}_0$ in Figure~\ref{fig:LameRS}. In terms of the basis functions \eqref{eq:solLame}, we can write this action as
\begin{equation}
    s^{\text{inst}}(u|\nu) = D^{\text{inst}}_{0} + D^{\text{inst}}_{1}G_1(u,u_0|\nu) + D^{\text{inst}}_{2}G_2(u,u_0|\nu) \quad.
\end{equation}
Similar to the classical action, we need three conditions to calculate the coefficients $D^{\text{inst}}_{k}$. To find these conditions, consider the properties of the action near the maximum of the potential:
\begin{enumerate}
    \item $s^{\text{inst}}(u|\nu)\to0$ as $u\to0\s$,
    \item $s^{\text{inst}}(u|\nu)$ is analytic near $u=0\s$,
    \item $\partial_u^{\s -}s^{\text{inst}}(u|\nu)\Bigl.\Bigr\rvert_{u=0}=-\dfrac{\pi \iu}{\sqrt{1-\nu}}\s$, where $\partial_u^{\s -}$ is the left derivative with respect to $u$.
\end{enumerate}
Akin to the classical action case, the first two conditions originate from the fact that $\mathcal{C}_0$ contracts to a point as $u\to0$. The third condition stems from a direct evaluation of the integral:
\begin{equation}
\begin{multlined}
    \partial_{u}^{\s-} s^{\text{inst}}(u|\nu)\Bigl.\Bigr\rvert_{u=0}
    = \partial_u^{\s-} \oint\limits_{\mathcal{C}_0} \sqrt{u + \cnt(x|\nu)} \d x \Bigl.\Bigr\rvert_{u=0}
    = \oint\limits_{\mathcal{C}_0} \dfrac{1}{2\sqrt{\cnt(x|\nu)}} \d x
    \\= 2\pi\iu\Res\left\{\dfrac{1}{2\s\cn(x|\nu)}, x=K(\nu)\right\} =- \frac{\pi \iu}{\sqrt{1-\nu}} \quad.
\end{multlined}
\end{equation}
We may freely choose the integration limit $u_0\s$. By setting $u_0=0$, we see that both functions $G_{1,2}(u,u_0)$ are zero at $u=0$, and the first condition renders us: $D^{\text{inst}}_{0}=0$. Since $G_2(u,0)$ is non-analytic near $u=0$, then $D^{\text{inst}}_{2}=0$. Calculation of the derivative of $G_1(u,0)$ gives:
\begin{equation}
    \partial_u^{\s-}G_1(u,0)\Bigl.\Bigr\rvert_{u=0} =+ \frac{\iu}{\sqrt{1-\nu}} \quad.
\end{equation}
From the third condition, it follows that
\begin{equation}
    \label{eq:instantonLame}
     s^{\text{inst}}(u|\nu) = -\pi \s G_2(u,0|\nu) = -\pi \int \limits_{0}^u \d v \dfrac{\lP_{-1/2}\left(\dfrac{(v+1)\nu-1-2v}{(v+1)\nu-1}\right)}{\sqrt{(v+1)\nu-1}}
    \quad.
\end{equation}

\begin{figure}[ht]
 \centering
 \includegraphics[width=0.8\textwidth]{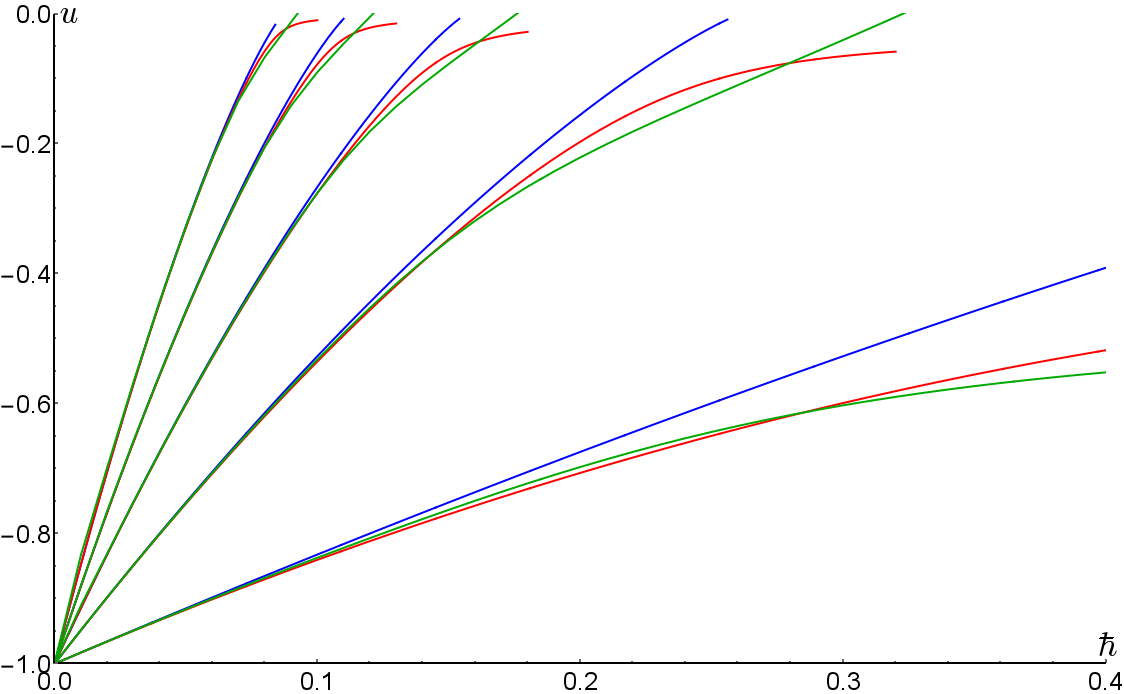}
 \caption{The five lowest energy states for the Lam\'e potential, which we calculated from the first-order (blue) and the second-order WKB approximation (red), compared to numerical solutions of the Schr\"odinger equation (green). Closer to the top of the potential $u=0$ and at larger values of $\hbar$, the corrections from the second term become more visible. Overall, the second-order results match the numerical calculation more closely.}
 \label{fig:specLame}
\end{figure}

For certain values of~$\nu$, the integrals in \eqref{eq:actionLame}, \eqref{eq:actionLame2} and \eqref{eq:instantonLame} can be expressed through special functions. However, the main advantage of these expressions over the contour integral is that they immediately produce the Taylor expansions in the energy $u$~--- at an arbitrary value of $u$. These expressions are the main result of this section, along with the Picard-Fuchs equation~\eqref{eq:PFLame}. The Bohr-Sommerfeld quantisation condition for the elliptic potential assumes the form
\begin{equation}
     s_c(u|\nu) = 2 \pi \hbar \left(n + \dfrac{1}{2}\right) \quad.
\end{equation}
To validate our results, we compare the ensuing energy levels with numerical solutions of the Schr\"odinger equation in Figure~\ref{fig:specLame}.

\section{Perturbative corrections to the Quantum Action Function}\label{sec:secondorder}

After a detailed discussion of the Picard-Fuchs method, we are in a position to apply similar ideas to calculate the second- and higher-order corrections in $\hbar$ in the generalised Bohr-Sommerfeld condition,
\begin{equation}\label{eq:generalizedBS}
 S(E) + \sum \limits_{l=2}^\infty \left(\dfrac{\hbar}{\iu}\right)^l \qactE_l(E) = S(E) + \sum \limits_{l=2}^{\infty} \left(\dfrac{\hbar}{\iu}\right)^l \oint \limits_{\mathcal{C}_R} \qmE_l(x,E) \d x = 2\pi\hbar \left(n + \dfrac{1}{2}\right) \quad.
\end{equation}
Here $\qactE_l(E)$ are the quantum corrections to the action function, while $\qmE_l(x,E)$ are corrections to the momentum. (For the rescaled values of these corrections, we shall use the notations $\qact_l$ and $\qm_l$, correspondingly.) For a full derivation of this condition from the WKB series and a recursive expression for $\qmE_l(x,E)$, we refer the reader to Appendix~\ref{app:generalizedBS}. Meanwhile, here we shall focus on the second-order correction given by
\begin{equation}\label{eq:secondcorr}
\begin{alignedat}{9}
 \qactE_2(E) =  \oint \limits_{\mathcal{C}_R} \qmE_2(x,E) \d x &= \oint \limits_{\mathcal{C}_R} \dfrac{P(x)P^{\s\prime\prime}(x)+P^{\s\prime}(x)^2}{24P(x)^3} \d x\\&= \oint \limits_{\mathcal{C}_R} \dfrac{-m\s V^{\s\prime\prime}(x)}{24\left[2m\left(E-V(x)\right)\right]^{3/2}} \d x \quad.
\end{alignedat}\end{equation}
Thus, the next-to-leading-order correction to the ordinary Bohr-Sommerfeld condition~\eqref{eq:BS} allows one to find the energy up to the second order in~$\hbar$.

\begin{sloppypar*}
The idea of our calculation takes its origin in an observation on the structure of the $1$\=/forms ${\qmE_l(x,E)\d x}$. Since these forms are obtained by differentiating the classical momentum $P(x, E)$, see equation~\eqref{eq:recursive}, they are defined on the same Riemann surface as the $1$\=/form ${\Lambda(E) = P(x, E)\d x}$. As was explained in detail in Section~\ref{sec:firstorder}, there exists only a finite number of linearly independent $1$\=/forms on a Riemann surface. Any other $1$\=/form can be expressed as a linear combination of these. Specifically, we can express a quantum correction $\qmE_l(x,E)\d x$ as a linear combination of the classical action $1$\=/form $\Lambda(E)$ and its derivatives $\Lambda_k(E)=\partial_E^{\s k}\Lambda(E)$, up to an exact form:
\end{sloppypar*}
\begin{equation}
 \qmE_l(x,E)\d x = \sum_{k=0}^K \gamma_k\Lambda_k(E) + \d f_2 \quad.
\end{equation}
Akin to the derivation of equation~\eqref{eq:ODE}, we can integrate this expression along $\mathcal{C}_R$ to obtain
\begin{equation}
 \qactE_l(E) = \sum_{k=0}^K \gamma_k\partial_E^{\s k}S(E) \quad.
\end{equation}
This formula implies that the quantum corrections to the classical action {\em at all orders} can be expressed through the derivatives of the classical action itself. While a similar fact is implied by equations \mbox{(2.12~--~2.13)} in \cite{Basar:2017hpr}, here we offer a different argument based on geometric consideration. Be mindful, though, that the coefficients $\gamma_k$ themselves may depend on the energy $E$.

Below we demonstrate how to obtain the second-order corrections for the sextic double-well and the Lam\'e potential. At this point, it would be in order to highlight that this calculation is similar to \textit{deriving} the Picard-Fuchs equation. However, it does not require \textit{solving} a differential equation and matching the correct boundary conditions. Calculating of higher-order WKB corrections is also a convenient way to generate the perturbative expansion. As an independent check of our results, in Appendix~\ref{app:recon} we shall obtain the first few terms of the perturbative expansion by inverting the generalised Bohr-Sommerfeld quantisation condition and solving it for the energy.

\subsection{The sextic double-well potential}

We start by calculating the correction of the order $\hbar^2$ to the classical action of the sextic potential~\eqref{eq:psextic}. Equation~\eqref{eq:secondcorr} entails:
\begin{equation}
    \qm_2 (y,u) \d y =
    \dfrac{1-15y^4}{24\left(\dfrac{2}{3^{3/2}}u + y^2 - y^6\right)^{3/2}} \d y
    \quad.
\end{equation}
From Section~\ref{sec:sextic}, we know that in the space of symmetric $1$\=/forms the basis can be chosen as $\{\lambda_k(u)\}_{k=0}^3\s$. Consequently,  the $1$\=/form $\qm_2(y,u)\d x$ can be expressed, up to an exact form, as their linear combination:
\begin{equation}
    \label{eq:sexticlincomb2}
    \qm_2(y,u)dy - \gamma_0\lambda_0(u) - \gamma_1\lambda_1(u) - \gamma_2\lambda_2(u) - \gamma_3\lambda_3(u) = \d f_2 \quad.
\end{equation}
The left-hand side of~\eqref{eq:sexticlincomb2} is a polynomial of degree $12$ in $y$, divided by $p(y,u)^5$. This suggests choosing the following expression for the exact form on the right-hand side of~\eqref{eq:sexticlincomb2}:
\begin{equation}
    \label{RHSjah}
    \d f_2 =
    \d{} \left[ \dfrac{R_7(y)}{p(y,u)^3} \right] \d y \quad,\qquad
    \text{where}\qquad
    R_7(y) = \sum \limits_{n=0}^7 b_n \s y^n \quad.
\end{equation}

We now substitute~\eqref{RHSjah} into~\eqref{eq:sexticlincomb2}, multiply both sides by $p(y,u)^5$, and then make sure that the coefficients in front of each power of $y$ vanish identically. This allows us to determine $b_n$, as well as the constants in~\eqref{eq:sexticlincomb2}:
\begin{equation}\begin{alignedat}{8}
    \gamma_0 &= 0 \quad&&,\qquad
    &&\gamma_1 &&= -\dfrac{45}{16}\s u\s \quad&&,\qquad\\
    \gamma_2 &= - \dfrac{9}{16} (-13+35\s u^2)\quad&&,\qquad
    &&\gamma_3 &&= -\dfrac{135}{16} (u^3-u)\quad&&.
\end{alignedat}\end{equation}
This way we get:
\begin{equation}\label{eq:sexticsecond}
    \qact_2(u) =
     -\dfrac{45}{16}\s u\s  s^{(1)}(u)
     - \dfrac{9}{16} (-13+35\s u^2)  s^{(2)}(u)
     -\dfrac{135}{16} (u^3-u)  s^{(3)}(u) \quad,
\end{equation}
$\s s(u)\s$ being the classical action from equation~\eqref{eq:actionsextic}. Switching back to the original variables gives us
\begin{equation}
    \qactE(E) = \dfrac{d}{2\s b^{\s 2}}
    \qact_2 (u)
    \quad,\qquad\text{where}\quad
    u =
    \sqrt{\dfrac{27 d}{4\s b^{\s 3}}} E
    \quad.
    \label{sigma2final}
\end{equation}

It does not hurt to reiterate the meaning of result~\eqref{eq:sexticsecond}: owing to the fact that the correction to the classical momentum $\qm_2(x,y)$ inhabits the same Riemann surface as the momentum $p(x, y)$ itself, we managed to express the correction to the classical action $\qact_2(u)$ through the action $s(u)$. It means that having calculated the classical action $s(u)$~--- by the method we used in Section~\ref{sec:sextic} or by any other method~---~we can efficiently calculate the higher-order WKB corrections to it, as demonstrated above. On each step, one needs to calculate $\s\qm_l(y,u)\s$ using the recursive relation given in Appendix~\ref{app:generalizedBS}, and to substitute it into equation~\eqref{eq:sexticlincomb2}. Then, one only needs to identify a generic expression for the exact form $\d f_l$, and to match the coefficients of the polynomials on the RHS and LHS, in order to find the coefficients $\{\gamma_k\}_{k=0}^3$.

With the second-order correction to the action at hand, we can now find the energy spectrum with an increased precision. This can be accomplished via using the formulae
\begin{subequations}\label{bzz}
\begin{alignat}{9}
     s_1(u_n) + \left(\dfrac{\hbar}{\iu} \right)^2 \qact_2(u_n) &= 2 \s \pi \s \hbar \s \left( n + \dfrac{1}{2} \right) \quad&&, \qquad -&&1&&<u&&<0 \quad&&,\\
     \tilde{s}_3(u_n) + \left(\dfrac{\hbar}{\iu} \right)^2 \tilde{\qact}_2(u_n)&= 2 \s \pi \s \hbar \s \left( n + \dfrac{1}{2} \right) \quad&&, \qquad &&0&&<u&&<1 \quad&&.
\end{alignat}
\end{subequations}
The results are shown in Figure~\ref{fig:sexticspec}.

\subsection{The elliptic potential}

Lastly, we calculate the correction of the order $\hbar^2$ to the classical action~\eqref{eq:actionLame} of the Lam\'e potential. In this case, equation~\eqref{eq:secondcorr} yields
\begin{equation}
    \qm_2(x,u|\nu) \d x=
    \dfrac{1-\nu + (4\s\nu-2) \cnt (x|\nu) - 3\s \nu\s \cntt (x|\nu) }{24\left(u + \cnt (x|\nu)\right)^{3/2}} \d x \quad.
\end{equation}
As discussed in Section~\ref{sec:Lame}, the set $\{\mu_k\}_{k=0}^2$ is a basis in the space of symmetric $1$\=/forms. So we can express $\qm_2 (x,u|\nu)\d x$, up to a total derivative, as
\begin{equation}
    \label{eq:Lamelincomb2}
    \qm_2 (x,u|\nu)\d x = \delta_0 \mu_0 (u|\nu) + \delta_1 \mu_1 (u|\nu) + \delta_2 \mu_2 (u|\nu) + \d g_2 \quad,
\end{equation}
The considered case is somewhat simpler as compared to the sextic potential, because one can match the coefficients $\delta_k$ in the above equation when setting the exact form equal to zero, $\d g_2=0$. Multiplied by $p(x,u|\nu)^3$, equation~\eqref{eq:Lamelincomb2} entails:
\begin{equation}\begin{multlined}
 \dfrac{1}{24} \left( 1-\nu + (4\s\nu-2) \cnt (x|\nu) - 3\s \nu\cntt (x|\nu) \right) =\\
 = -\dfrac{\delta_2}{4} + \dfrac{\delta_1}{2}u + \delta_0\s u^2 + \left(\dfrac{\delta_1}{2} +2\s\delta_0\s u \right)\cnt(x|\nu) + \delta_0 \cntt(x|\nu)\quad.
\end{multlined}\end{equation}
Equating the coefficients next to the powers of $\s\cnt(x|\nu)\s$ results in
\begin{equation}\begin{alignedat}{8}
    \delta_0 &= - \dfrac{\nu}{8} \quad,\qquad
    \delta_1 = \dfrac{1}{6}(2\nu - 1 + 3\s\nu\s u)\quad&&,\\
    \delta_2 &= \dfrac{1}{6} \left(\nu-1 + 2(2\s\nu-1)u + 3\s\nu\s u^2\right) \quad&&,
\end{alignedat}\end{equation}
which in turn yields
\begin{equation}\begin{multlined}
    \qact_2(u|\nu) = - \dfrac{\nu}{8}  s (u|\nu) + \dfrac{1}{6}(2\s\nu - 1 + 3\s\nu \s u)  s^{(1)} (u|\nu)
    \\+\dfrac{1}{6} \left(\nu-1 + 2(2\nu-1)u + 3\s\nu\s u^2\right)  s^{(2)} (u|\nu) \quad.
\end{multlined}\end{equation}
Hence, in just a few steps, we have found an expression for the second-order quantum corrections in terms of only the classical action and its first two derivatives. Applying this to the second-order Bohr-Sommerfeld quantisation condition,
\begin{equation}
     s(u|\nu)  + \left(\dfrac{\hbar}{\iu} \right)^2 \qact_{2}(u_n|\nu) = 2 \s \pi \s \hbar \s \left( n + \dfrac{1}{2} \right) \quad,
\end{equation}
we arrive at the results that are shown in Figure~\ref{fig:specLame}. In terms of the original variables, the correction is
\begin{equation}
    \label{sigmatla}
    \qactE_2(E|\nu) = \dfrac{1}{a}\s\qact_2 (u|\nu)
    \quad,\qquad
    \text{where}
    \quad
    u = \dfrac{E-b}{a\s\nu} -1 \quad.
\end{equation}

\section{Summary}\label{sec:summary}

In this paper, we have demonstrated how arguments from algebraic topology can be used to perform calculations efficiently in classical and quantum mechanics. The main objects of our studies are the classical (abbreviated) action and the quantum mechanical corrections to it, which are derived from the WKB series of the quantum action function. By continuation to complex coordinates, these quantities can be expressed in terms of integrals along closed contours on the Riemann surface of the classical momentum. For the Lam\'e and sextic double-well potentials, we have shown in detail how to calculate such integrals in a short and elegant manner, our results being similar to those obtained in~\cite{Basar:2017hpr} for genus-1 potentials. These two potentials were chosen by us because they show up in the studies of quasi-exactly solvable models~\cite{ShifmanTurbiner,Shifman,Dual,LONG}, the underlying Riemann surface is of genus 2, and also because they represent the two main types of spectra in quantum mechanics~--- bound states in an unbounded potential and a band/gap structure in a continuous periodic potential. Their actions exhibit a duality property~\cite{LONG} which we use to check our results.

We first consider the classical (abbreviated) action which, besides its importance in classical mechanics, is used in the Bohr-Sommerfeld quantisation condition to approximate quantum-mechanical energy levels. We demonstrate how to relate this action to an integral on a complex manifold, and show how this integral is linked to an ordinary differential equation named the \textit{Picard-Fuchs equation}.
We elucidate, step by step, how the topological properties of the Riemann surface and the analytic properties of the integrand can be utilised to derive the Picard-Fuchs equation. We discuss the differences in constructing the Riemann surfaces for the two cases of an unbounded and a periodic potential; and we also point out that the resulting manifolds turn out to be topolopgically equivalent. Furthermore, we show a straightforward recipe for deriving the Picard-Fuchs equation, and explain how to obtain the coefficients linking the action to its basis solutions. From there, we calculate the energy levels at order~$\hbar$ via the Bohr-Sommerfeld quantisation rule, and show in Appendix~\ref{app:above} the analogous calculation for the classical action above the maxima of the potentials.

Building on these results, we consider the perturbative calculation of the quantum mechanical analogue of the abbreviated action, a \textit{quantum action function} showing up in the generalised Bohr-Sommerfeld quantisation condition. We argue that \textit{all} the perturbative corrections to the quantum momentum function are integrals defined on the same Riemann surface as the classical action. Following the same arguments as in the derivation of the Picard-Fuchs equation, we show that these quantum corrections, at all orders in $\hbar$, can be expressed through the action and its first few derivatives. Here the maximum number of derivatives is by one smaller than the degree of the Picard-Fuchs equation. We explicitly calculate the corrections of order $\hbar^2$ for the sextic double-well and Lam\'e potentials. As an independent check of our results, we compare those corrections to perturbative expansions of the generalised Bohr-Sommerfeld condition in Appendix~\ref{app:recon}. We want to emphasize that calculating the corrections is equivalent to deriving the Picard-Fuchs equation, though it requires neither solving a differential equation nor finding boundary conditions. So we acquire a computationally simple method to calculate quantum corrections to the classical action. These permit to obtain improved approximations to the quantum energy levels.

\section{Acknowledgements}

The authors are deeply thankful to Peter Koroteev for introducing to them the concepts of algebraic topology necessary for this work. \mbox{T.G.} is grateful to Michael Janas and Alex Kamenev for many helpful discussions pertinent to this project. \mbox{T.G.} was supported in part at the Technion by an Aly Kaufman fellowship.
\mbox{M.K.} is deeply grateful to Michael Efroimsky for meticulous reading of the manuscript.

\newpage
\begin{appendices}

\section{{Basics of algebraic topology} \label{centrid}}\label{app:topology}

The tools from algebraic topology, which we used to derive the Picard-Fuchs equation, can also serve many other purposes in physics. Therefore we introduce these basic concepts here (with the focus on complex manifolds), and show the derivation of the Picard-Fuchs equation~--- this time without referring to the classical action, which is just one of the possible applications of this concept.

In physics, one often has to deal with functions defined as integrals whose free parameter is the functions' argument:
\begin{equation}
    F(u) = \oint \limits_{\mathcal{C}_0} \integrand(x,u)\d x = \oint \limits_{\mathcal{C}_0} \integrandform(u) \quad.
    \label{F}
\end{equation}
In many such cases the integrals can only be evaluated analytically for specific values of the parameter. For this reason, special functions that are given by differential equations are commonly defined through their integral representation. This allows one to study the analytic properties of the function in various domains of the parameter's values.

In the majority of situations arising in quantum mechanics and quantum field theory, $F(u)$ is not a well-studied special function. However, suppose that a differential equation obeyed by $F(u)$ is known. Then, by matching the boundary conditions, one may be able to express $F(u)$ through the basis of solutions of the differential equation, which are special functions with well-known properties. Additionally, a differential equation for $F(u)$ is of more use than the integral form~\eqref{F}, when it is necessary to study the asymptotics of $F(u)$. The Picard-Fuchs method implements this approach, constructing a differential equation for a known integral form. In a sense, this procedure is inverse to solving the differential equation.

We begin by extending the integrand in~\eqref{F} to the complex plane, and considering integration along a closed contour (cycle). In physically relevant cases, $\integrand(x,u)$ is not a globally defined analytic function. However, one can define a complex manifold $\mathcal{M}$ on which $\integrand(x,u)$ is globally analytic~--- the Riemann surface of $\integrand(x,u)$. The Picard-Fuchs approach to deriving a differential equation for the function $F(u)$ is based on studying the global properties of $\mathcal{M}$. Topologically, this manifold is equivalent to a multi-torus, whose number of holes is referred to as the genus $g$ of the manifold. In distinction from the complex plane, on a multi-torus there exist cycles which cannot be continuously deformed to a point, namely those encircling a handle or a hole (see also Figure~\ref{fig:doubletorus}).

While in the complex plane the integrand $\integrand(x,u)$ is a globally multi-valued function, it is locally single-valued almost everywhere. The exceptions are the \textit{branch points}. Performing analytic continuation to the entire complex plane, one encounters the lines of discontinuity named \textit{branch cuts}. These lines may be chosen arbitrarily, though must always connect pairs of branch points. Each of the multiple values of $\integrand(x,u)$ defines a copy of the complex plane. Together, these multiple sheets form the Riemann surface of the function $\integrand(x,u)$, whereon this function is analytic everywhere.

Typically, on a complex $1$-dimensional manifold we make no distinction between cycles which can be continuously deformed into one another (i.e., are homotopically equivalent), since the integrals of any analytic functions along these cycles coincide. The reason is that any two such cycles differ by a cycle which is a boundary of some region $\mathcal{K}$ of the surface, along which an integral of any analytic function vanishes:\s\footnote{~By Stokes's theorem, the integral along the boundary cycle is equal to the area integral over its differential, $\displaystyle\int \limits_{\partial\mathcal{K}} \exampleform = \int \limits_{\mathcal{K}} \d\exampleform$. On a 1-dimensional manifold, the differential of any 1-form vanishes, $\d\exampleform=0$.}
\begin{gather}
    \mathcal{C}_1 \sim \mathcal{C}_2 \qquad\Longleftrightarrow\qquad
    \mathcal{C}_1 - \mathcal{C}_2 = \partial \mathcal{K}\quad,\\
    \forall \exampleform: \quad
    \int \limits_{\mathcal{C}_1} \exampleform =
    \int \limits_{\mathcal{C}_2} \exampleform
    + \int \limits_{\partial\mathcal{K}} \exampleform
    = \int \limits_{\mathcal{C}_2} \exampleform
    \quad.
\end{gather}
In other words, we consider only the equivalence classes of such cycles. This serves as a motivation for defining the first homology group $H_1(\mathcal{M})$ of the manifold, which is the group of equivalence classes of cycles modulo boundaries. The group operation is the merging of two cycles, while the inversion operation is changing the orientation of a cycle.

For a complex 1-dimensional manifold of finite genus $g$, there exists a finite basis of cycles ${\{\mathcal{C}_k\}_{k=1}^N}\;$:
\begin{subequations}
\begin{gather}
    \sum\limits_{k=1}^{N} a_k\s \mathcal{C}_k = \partial\mathcal{K}
    \qquad \Longrightarrow \qquad a_k=0\quad, \label{clin1}
    \\
    \forall\s\mathcal{C}_0 \ \: \exists \{a_k\}_{k=1}^N:\quad
    \mathcal{C}_0 = \sum \limits_{k=1}^{N} a_k \s \mathcal{C}_k + \partial\mathcal{K}
    \quad,\label{clin2}
\end{gather}
\label{czhopa}
\end{subequations}
where $\mathcal{C}_0$ is an arbitrary cycle in $\mathcal{M}$ and $a_k$ are integers, while $\partial \mathcal{K}$ is the boundary of a closed region. Thereby, $H_1(\mathcal{M})$ is a $\mathbb{Z}$-module, a structure similar to a vector space in which the scalars are taken from a ring instead of a field. The total number of independent cycles is ${N=2g}$ for a manifold without singularities, and ${N=2g+s-1}$ for a manifold with $s$ punctured points.

Similar is the situation with $1$-forms defined on $\mathcal{M}$. Integration of an exact form (a total derivative of an analytic function, $\d f=\partial_x f(x)\d x$) over \textit{any} cycle gives zero.\s\footnote{~This is again by Stokes' theorem, $\displaystyle\int \limits_{\mathcal{C}} \d f = \int \limits_{\partial\mathcal{C}} f$. When a contour is closed, its boundary is zero, $\partial\mathcal{C}=0$.} Accordingly, two $1$-forms which differ by an exact $1$-form are indistinguishable upon integration along any closed contour:
\begin{gather}
    \exampleform_1 \sim \exampleform_2 \qquad\Longleftrightarrow\qquad
    \exampleform_1 - \exampleform_2 = \d f\quad,\\
    \forall\s \mathcal{C}_0: \quad
    \int \limits_{\mathcal{C}_0} \exampleform_1 =
    \int \limits_{\mathcal{C}_0} \exampleform_2
    + \int \limits_{\mathcal{C}_0} \d f
    = \int \limits_{\mathcal{C}_0} \exampleform_2
    \quad.
\end{gather}
This defines an equivalence class of $1$-forms, the first cohomology group $H^1(\mathcal{M})$. The said group is also a vector space over complex numbers. In it, one can define a basis $\{\exampleform_n\}_{n=1}^N$:
\begin{subequations}
\begin{gather}
    \sum\limits_{n=1}^{N} b_n \s\exampleform_n = \d f
    \qquad \Longrightarrow \qquad b_n=0\quad, \label{lin3}
    \\
    \forall\s\exampleform_0 \ \: \exists \{b_n\}_{n=1}^N,\,\d f:\quad
    \exampleform_0 = \sum \limits_{n=1}^{N} b_n \s \exampleform_n + \d f \label{lin4} \quad,
\end{gather}
\end{subequations}
where $ \exampleform_0$ is an arbitrary 1-form defined on $\mathcal{M}$ and $\d f$ an exact 1-form.

An important result from topology, whereon our further discussion will rely, is that for an oriented 1-dimensional complex manifold $\mathcal{M}$ (possibly with a finite number of punctured points) the dimensions of the first homology and first cohomology groups are equal:\s\footnote{~
At this point, one may be tempted to refer to the Poincar\'e  duality. This duality, however, does not hold for manifolds with punctured points. Fortunately, the weaker result~\eqref{hhdu} is sufficient for our needs.}
 \begin{equation}
 \label{hhdu}
    \dim_\mathbb{Z} H_1(\mathcal{M}) = \dim_\mathbb{C} H^1(\mathcal{M}) \quad.
 \end{equation}

In application to our problem, this relation implies that
\textit{the number} $N$ \textit{of linearly independent cycles (modulo boundary) on a given differential complex manifold} $\s \mathcal{M}\s $ \textit{is equal to the number of the linearly independent $\s 1$-forms (modulo an exact $1$-form)}.

As a corollary, we can define the \textit{de Rham basis}. For a basis of cycles $\{\mathcal{C}_k\}_{k=1}^N\;$, there exists a basis of 1-forms $\{\exampleform_n\}_{n=1}^N$ such that\s\footnote{~Note that in the main text of our paper we mostly discuss a different basis, the one obtained by differentiating a certain $1$-form with respect to its parameter, see equation~\eqref{difbas}. This is not the de Rham basis.}
\begin{equation}
    \oint_{\mathcal{C}_k}\exampleform_n = \delta_{k,n} \quad.
\end{equation}

\begin{sloppypar*}
We are now in a position to describe the Picard-Fuchs method of constructing the differential equation for a function $F(u)$ defined by~\eqref{F}. We start out with a key observation that taking a derivative of a 1-form ${\integrandform(u) = \integrand(x,u)\d x}$ with respect to the parameter $u$ does not lead us away from the manifold $\mathcal{M}$. In other words, every differentiation produces another 1-form defined on the same manifold $\mathcal{M}$. Calculating the first $N$ derivatives of~$\integrandform(u)$,
\end{sloppypar*}
\begin{equation}\label{difbas}
    \integrandform_k(u)
    \equiv \partial^{\s k}_{u} \integrandform(u)
    \qquad,\quad
    k=0\ldots N
    \quad,
\end{equation}
we obtain a set $\;\{\integrandform_k(u)\}_{k=0}^N\;$ of $\;(N+1)\;$ $1$-forms (including the original form), of which \textit{at most} $N$ are linearly independent. Consequently, we can write
\begin{equation}
    \sum_{k=0}^K \alpha_k \s \integrandform_k(u) = \d f 
    \label{eq:app:lincomb}
\end{equation}
for a non-trivial set $\{\alpha_k\}$ and an exact form $\d f$. Note that $K$ must be less or equal to $N$, for the entire space of 1-forms is not necessarily spanned by the derivatives of $\integrandform(u)$. The examples in the main text visualise this effect. Upon integrating along the cycle $\mathcal{C}_0$ in \eqref{F}, the last equation turns into:
\begin{equation}
    \oint_{\mathcal{C}_0} \sum_{k=0}^K \alpha_k \s \integrandform_k(u) =
    \sum_{k=0}^K \alpha_k \s
    \partial_u^{\s k}F(u) = 0 \quad.
\end{equation}
Hence the linear combination \eqref{eq:app:lincomb} turns into a differential equation for the function $F(u)$, the Picard-Fuchs equation. In our paper, we use this equation to find the classical action ${s(u)}$ which is  obtained via integrating the classical momentum $1$\=/form ${p(x,u) \d x}$.

This discussion suggests the following method of constructing a differential equation for the function $F(u)$:
\begin{enumerate}
    \item Investigate the global properties of the Riemann surface $\s \mathcal{M}\s $, to determine the number $N$ of linearly independent integration cycles, which is the same as the maximum number of linearly independent 1-forms available on $\s \mathcal{M}\s $.
    \item Evaluate the first $\s N\s $ derivatives of the 1-form $\integrandform(u)$. Since integrating with respect to the coordinate $x$ commutes with taking a derivative with respect to the parameter $u$ on the RHS of~\eqref{F}, we conclude that
    \begin{equation}
        \partial_u^{\s k}F^{(k)}(u) = \oint \limits_{\mathcal{C}_0} \partial_u^{\s k}\integrand(x,u) \d x = \oint \limits_{\mathcal{C}_0}  \integrandform_k(u) \quad.
    \end{equation}
    \item Analyse the global properties of the 1-form $\integrandform(u)$ and its derivatives, and determine whether all the basis 1-forms can be expressed in terms of those. If a basis contains $L$ 1-forms that can not be expressed in terms of the derivatives, then $K=N-L$.
    \item Construct a condition of these $\s 1$-forms' linear dependence. To this end, find such coefficients $\s \alpha_k\s $ on the LHS of~\eqref{eq:app:lincomb} that the expression on the RHS is a total derivative. Be mindful that the coefficients $\s \alpha_k\s $ are allowed to depend on $u$ but not on $x$:
    \begin{equation}
        \alpha_0(u)\s \integrandform_0(u) + \alpha_1(u)\s \integrandform_1(u) + \ldots + \alpha_K(u)\s  \integrandform_K(u) = \d f \quad.
        \label{eqraw}
    \end{equation}
    \item Notice that, after being integrated over the contour $\mathcal{C}_0$\s, equation \eqref{eqraw} turns into
    \begin{equation}
        \alpha_0(u)\s F(u) + \alpha_2(u)\s F^{(1)}(u) + \ldots
        + \alpha_K(u)\s F^{(K)} (u) = 0 \quad,
        \label{PFeq}
    \end{equation}
    which is the desired Picard-Fuchs equation for the function $F(u)$.
\end{enumerate}

\section{Generalised Bohr-Sommerfeld quantisation condition}\label{app:generalizedBS}

Here we provide a squeezed inventory of the facts from Quantum Mechanics, which are used in our study.  Our starting point is the Schr\"odinger equation in one dimension:

\begin{equation}
    \widehat{H} \s \psi(x) = E \s \psi (x)
    \quad,\qquad
    \widehat{H} = \dfrac{\widehat{P}^{\s 2}}{2\s m} + V(x)
    \quad.
    \label{Sch}
\end{equation}
Performing the substitution
\begin{equation}
    \psi(x) = \exp\left( \iu\s\qactE(x, E) / \hbar \right) \quad,\label{ch}
\end{equation}
we observe that the function $\qactE^{\s\prime}(x,E) \equiv \partial_x \qactE(x,E)$ satisfies the Riccati equation:
\begin{equation}
    (\qactE^{\s\prime}(x, E))^2 + \dfrac{\hbar}{\iu} \qactE^{\s\prime\prime}(x, E) = 2\s m\s(E - V(x)) \quad.\label{ric1}
\end{equation}
It ensues from this equation that in the limit of $\hbar \to 0$ the function \begin{equation}
    \qmE(x, E) \equiv  \qactE^{\s\prime}(x, E)\quad.
    \label{290}
\end{equation}
satisfies the equation for the classical momentum. So it may be termed as the \textit{quantum momentum function} (QMF).

Accordingly, the equation~\eqref{ric1} takes the form:
\begin{equation}
    \qmE^{\s 2}(x, E) + \dfrac{\hbar}{\iu}  \qmE^{\s\prime}(x, E) = 2\s m\s(E - V(x)) \quad.\label{ric2}
\end{equation}

The quantisation condition, whence the $n$-th energy level is determined, is normally obtained from the requirement of single-valuedness of the function $\psi(x)$. However, in~\cite{HamJa} a more interesting option was proposed. It was based on the fact that the wave function corresponding to the $n$-th energy level has $n$ zeros on the real axis, between the classical turning points (the latter points being the zeros of the classical momentum)~\cite{LL}.

In these zeroes, the QMF  has poles. Indeed, it trivially follows from (\ref{ch}) and (\ref{290}) that
\begin{equation}\label{pivoisemki}
    \qmE(x, E) = \dfrac{\hbar}{\iu} \dfrac{1}{\psi(x)} \dfrac{\d \psi(x)}{\d x}\quad.
\end{equation}
For analytic potentials, the pole of the function $\qmE(x,E)$ is of the first order, and the residue at this pole is $(-\iu \hbar)$. Therefore, the integral of the QMF along the contour $\mathcal{C}_R$ enclosing classical turning points is:
\begin{equation}
    B(E) = \oint \limits_{\mathcal{C}_R} \qmE(x, E) \s\d x = 2 \s \pi \s n\s\hbar\quad,\label{GBS}
\end{equation}
where the contour $\mathcal{C}_R$ should be close enough to the real axis, in order to avoid containing the poles and branch cuts of $\qmE(x,E)$, that are off the real axis. In the classical limit ($\hbar \to 0$), the series of poles inside $\mathcal{C}_R$ coalesces into a branch cut of the classical momentum~\cite{LP}.

The function $B(E)$ is sometimes referred to as the \textit{quantum action function} (QAF)~\mbox{\cite{LP,LP2}}, and the equality~\eqref{GBS} itself~---~as the generalised Bohr-Sommereld quantisation condition (GBS). The GBS is often employed as a starting point in studies of the spectra of quantum systems. It contains the same amount of information about the physical system as the original Schr\"odinger equation~\eqref{Sch}.

In the cases where the energy is sought in the form of an expansion over a small parameter, one typically distinguishes between the perturbative and non-perturbative kinds of contributions. From now on, we shall focus on the former kind. To do so, we shall employ the expansion of the QMF in the powers of $\hbar$ (the WKB method):
\begin{equation}
    \qmE(x,E) = \sum \limits_{k=0}^{\infty} \left(\dfrac{\hbar}{\iu}\right)^k \qmE_k(x,E) \quad.
    \label{L8}
\end{equation}


Substituting~\eqref{L8} into the Riccati equation~\eqref{ric2} gives a recursive relation
\begin{gather}
 \label{eq:recursive}
    \sum \limits_{l=0}^{k} \qmE_l(x,E)\s \qmE_{k-l}(x,E) + \qmE_{k-1}'(x,E) = 0 \quad,
\end{gather}
which allows us to express all the higher terms~\eqref{L8} through the classical momentum:
\begin{gather}\begin{gathered}
    \qmE_k(x,E) = - \dfrac{1}{2\s \qmE_0(x,E)} \left(\qmE^{\s\prime}_{k-1}(x,E)+\sum \limits_{l=1}^{k-1} \qmE_l(x,E) \s \qmE_{k-l}(x,E) \right)\quad,\\
    \qmE_0 (x,E) = P(x, E)\quad.
    \label{qmkp}
\end{gathered}\end{gather}

We define the $k$-th correction to the classical action as
\begin{equation}
 \qactE_k (E) =
 \int \limits_{\mathcal{C}_R} \qmE_k(x,E) \d x \quad.
\end{equation}
The series expansion in powers of $\hbar$ for the quantum action takes the form of
\begin{equation}
    \label{puffyvgn}
B(E) = S(E)
+ \dfrac{\hbar}{\iu} \qactE_1 (E)
+ \left(\dfrac{\hbar}{\iu}\right)^2 \qactE_2 (E) + \ldots \quad.
\end{equation}

Next, we substitute the expansion~\eqref{L8} into~\eqref{GBS} and obtain:
\footnote{~As we have already mentioned, the form of the equation above implies the neglect of the tunneling effects.
}
\begin{equation}
    B(E) = S(E) + \sum \limits_{k=1}^\infty
    \left(\dfrac{\hbar}{\iu}\right)^k
    \qactE_k(E)
    =2\s\pi\s n\s\hbar \quad.
    \label{GBSser}
\end{equation}

The zeroth and first terms in~\eqref{GBSser} render:
\begin{equation}
    S(E) + \dfrac{\hbar}{\iu}\s2\s\pi\s\iu\left(-\dfrac{1}{2}\right)
    = S(E) - \pi\s\hbar
    = 2\s\pi\s n\s\hbar \quad,
    \label{derBS}
\end{equation}

The constant arising from the first term is often referred to as Maslov index and can be calculated in various ways~\cite{LL}. Importantly, it does not depend on the form of the potential well. After moving it to the RHS of~\eqref{derBS}, we arrive at the famous Bohr-Sommerfeld quantisation condition:
\begin{equation}
    S(E) = 2\s\pi\s\hbar \left(n+\dfrac{1}{2}\right)\quad.
    \label{BS}
\end{equation}

One may also proceed with calculating the higher-order terms on the LHS of~\eqref{GBSser}. This will, for example, provide a way to generate the perturbative expansion in the cases where it exists. To this end, one will have to solve~\eqref{GBSser} for the energy, inverting the series term by term.
When evaluating the integrals, one should take into account that all the odd terms in the expansion of the QMF, starting from $k=3$, are total derivatives, so the corresponding integrals in~\eqref{GBSser} vanish.

\section{Classical action above the maximum}\label{app:above}
In this section, we calculate the classical action above the maximum at $u=0$ for the sextic and Lam\'e potentials.

\subsection{Above the local maximum in the sextic potential}

\begin{figure}[ht]
 \centering
 \includegraphics[width=0.45\textwidth]{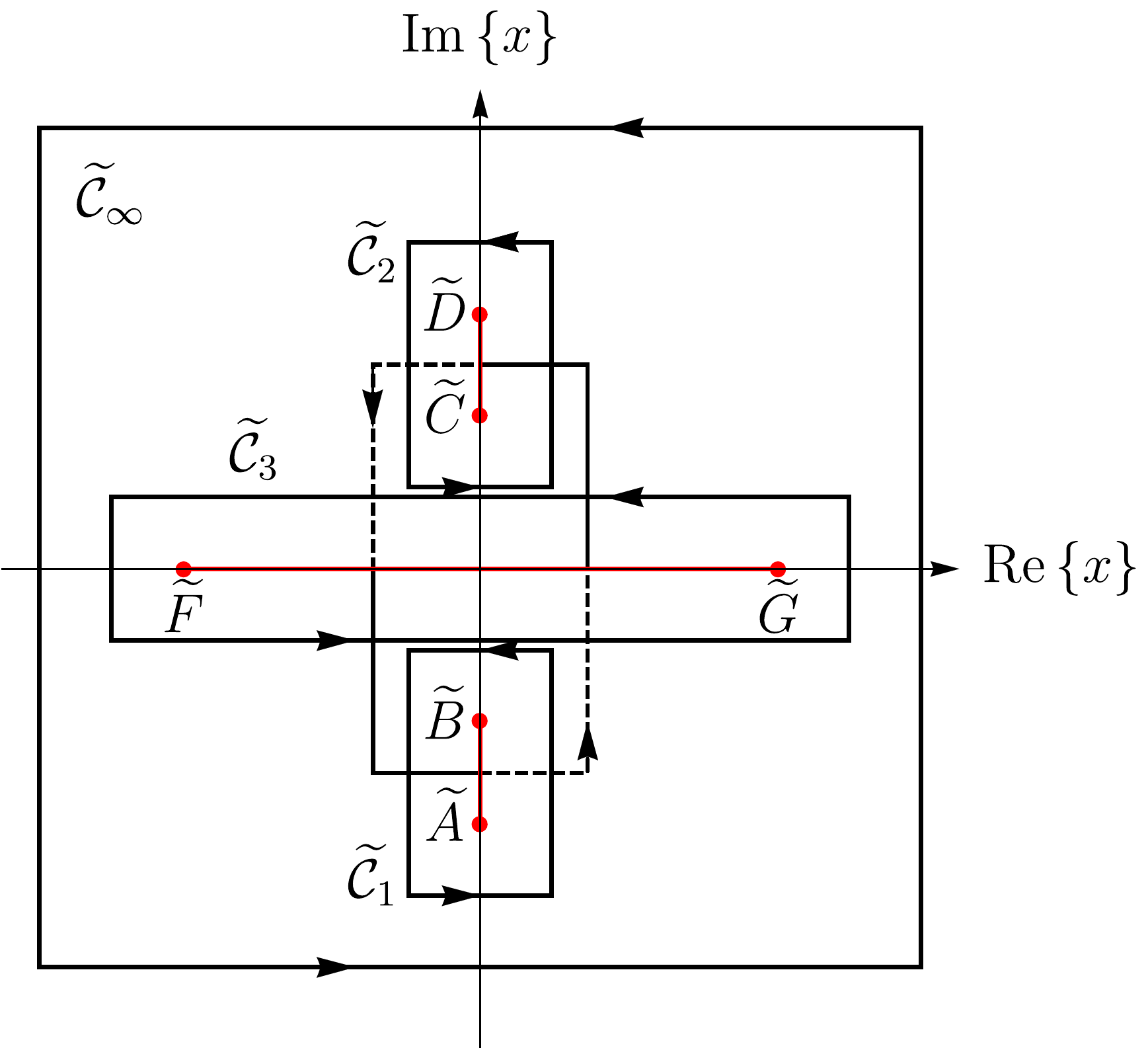}
 \caption{The integration cycles for $0 < E < -V_{\text{min}}$.}\label{fig:roots62}
\end{figure}

Consider the classical action above the local maximum at $u=0$. We use the same rescaled coordinates as those introduced in equation~\eqref{eq:psextic}, so we work in the regime $0<u<1$. Figure~\ref{fig:roots62} shows the integration cycles for this case. We define the periods as
\begin{equation}\label{eq:periodssextic2}
     \tilde{s}_j(u) = \oint \limits_{\widetilde{\mathcal{C}}_j} p(x,\s u)\d x \equiv \oint \limits_{\widetilde{\mathcal{C}}_j} \lambda(u) \quad,\qquad j=1,2,3,\infty\quad.
\end{equation}
For motion between the turning points $\widetilde{F}$ and $\widetilde{G}$, the classical action is calculated by integration over the cycle $\widetilde{\mathcal{C}}_3$. As before, we begin with investigating an auxiliary cycle~$\widetilde{\mathcal{C}}_2$ which encloses the points~$\widetilde{B}$ and~$\widetilde{C}$. Similarly to the previous case, this cycle shrinks to a point as~$u\to0$. So, in this limit, the integral over this cycle approaches zero, ${ \tilde{s}_2(0)=0}$, and is analytic in a vicinity of this point. Of the basis functions $F_k(u)$ defined by equation \eqref{eq:Solsextic}, only the functions $F_1(u)$ and $F_2(u)$ are analytic, while $F_0(u)$ and $F_3(u)$ are not. So the latter two functions cannot contribute to $\tilde{s}(u)$, whence $\widetilde{C}_{2,0}=\widetilde{C}_{2,3}=0$. To identify the other two coefficients, we again expand the integrand to the second order in $u$ and perform a residue calculation for both terms. This yields:
\begin{equation}
 \tilde{s}_2 (u) = \dfrac{2\pi \iu}{3^{3/2}}u + \mathcal{O}(u^3) \quad.
\end{equation}
Comparing this with the expansions of the two remaining basis solutions, $F_1(u)=u+\mathcal{O}(u^3)$ and $F_2(u)=u^2+\mathcal{O}(u^4)$, we identify the coefficients as $\widetilde{C}_{2,1}=\dfrac{2\pi \iu}{3^{3/2}}$ and $\widetilde{C}_{2,2}=0$. The action is, therefore,
\begin{equation}
    \tilde{s}_2 (u) = \dfrac{2\pi \iu}{3^{3/2}}F_1(u).
    \label{eq:sexticS0prime}
\end{equation}
This action does not carry much physics with it, but we shall need this result at the next step of our calculation, as we turn to
$\tilde{s}_3(u)$.

\begin{sloppypar*}
Near $u=0$, we perform a monodromy transformation similar to that in Figure~\ref{fig:monodromy}, ${u\to u\s \me^{2\pi \iu}}$. It transforms the cycle as ${{\widetilde{\mathcal{C}}_3\to \widetilde{\mathcal{C}}_3+2\s\widetilde{\mathcal{C}}_2}}$. With every such monodromy transformation, the action ${\tilde{s}_3(u)}$ obtains an additional contribution of ${2\tilde{s}_2(u)}$, which allows us to write:
\end{sloppypar*}
\begin{equation}
    \tilde{s}_3(u) = \widetilde{Q}_3(u) + 2\dfrac{\tilde{s}_2(u)}{2 \pi \iu}\log(u) \quad,
\end{equation}
with the function $\tilde{Q}_3(u)$ being  analytic near $u=0$. The only non-analyticity comes from $F_3(u)$, and we find the corresponding coefficient to be
\begin{equation}
   \label{eq:sexticC33}
    \widetilde{C}_{3,3}=-\left[6\s\Gamma\left(\dfrac{1}{3}\right)\Gamma\left(\dfrac{1}{6}\right)\right]^{-1}.
\end{equation}
Also, in the limit of $u\to0^{\s+}$ the integral can be evaluated analytically: $\tilde{s}_3(0)=\dfrac{\pi}{2}$. Besides $F_3(u)$, the only function nonvanishing in $u=0$ is $F_0(u)$. From $F_3(0)=\pi\Gamma\left(\dfrac{1}{3}\right)\Gamma\left(\dfrac{1}{6}\right)$ and $F_0(0)=1$, we obtain:
\begin{equation}
    \widetilde{C}_{3,0}=\dfrac{\pi}{3}\quad.
    \label{eq:sexticC30}
\end{equation}

Lastly, we consider the behaviour near $u=1$. In the sense of the structure of the branch cuts, this is not a special value for $\widetilde{\mathcal{C}}_3$, so the resulting integral $\tilde{s}_3(u)$ is analytic in this point. However, the two basis functions $F_1(u)$ and $F_2(u)$ have logarithmic non-analyticities. This means that they have to cancel, which yields a condition on the coefficients. We also can evaluate the integral analytically at $u=1$: $\tilde{s}_3(1)=\pi$. This gives a second constraint on the two remaining coefficients, which uniquely defines them as
\begin{equation}
   \label{eq:sexticC31C32}
   \widetilde{C}_{3,1} = \widetilde{C}_{3,2} = 0 \quad.
\end{equation}
The coefficients in equations~\mbox{(\ref{eq:sexticC33}-\ref{eq:sexticC31C32})} fully define the classical action above the local maximum of the double-well potential:
\begin{equation}\label{eq:actionsexticabove}
    \tilde{s}_3(u) = \frac{\pi}{3} - \left[6\s\Gamma\left(\dfrac{1}{3}\right)\Gamma\left(\dfrac{1}{6}\right)\right]^{-1} F_3(u)\quad.
\end{equation}




\subsection{Above the local maximum in the periodic potential}

\begin{figure}[ht]
 \centering
 \includegraphics[width=0.7\textwidth]{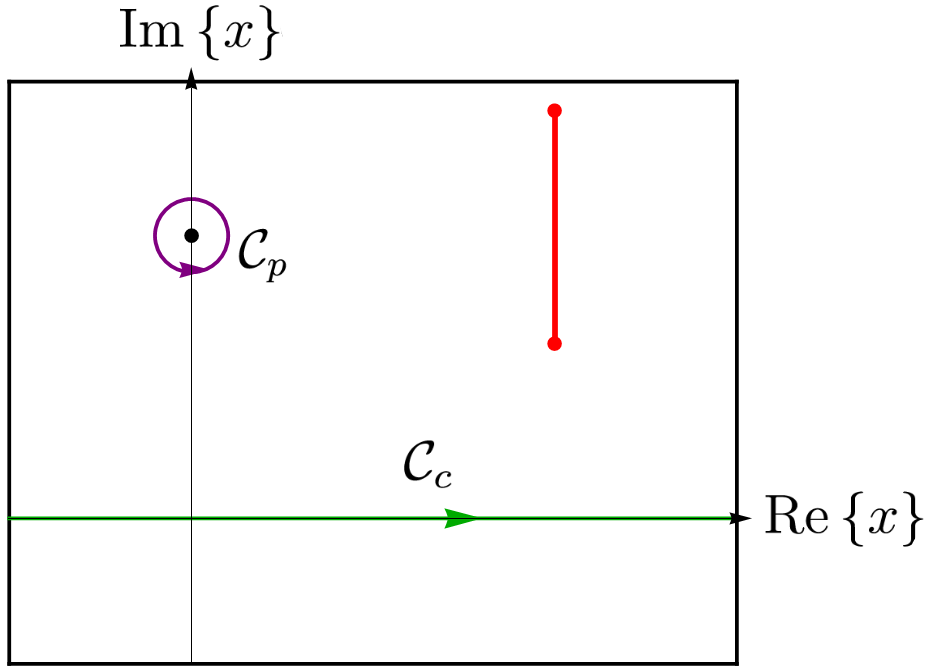}
 \caption{The structure of the fundamental parallelogram for the classical momentum for the Lam\'e potential, for energies $u>0$. The branch cut (red) runs in the imaginary direction and does not intersect the cycle $\widetilde{\mathcal{C}}_c$ (green) which runs along the real axis and corresponds to classical motion. This cycle is closed by periodicity.}
 \label{fig:Lameabove}
\end{figure}

Here we calculate the classical action for quasi-free motion in the periodic potential~\eqref{eq:Ve}. We write the periods as generic linear combinations of solutions~\eqref{eq:solLame} of the Picard-Fuchs equation
\begin{equation}\label{eq:Lameabove}
 \tilde{s}(u|\nu) = \widetilde{D}_0G_0 + \widetilde{D}_1G_1(u,u_0|\nu) + \widetilde{D}_2G_2(u,u_0|\nu) \quad.
\end{equation}
The relevant  integration cycle is shown in Figure~\ref{fig:Lameabove}. One full period of classical motion above the potential corresponds to moving once through the unit cell. Thence the classical action is
\begin{equation}\label{eq:SLameabove}
 \tilde{s}(u|\nu) = \int \limits_{-K(\nu)}^{K(\nu)} p(x,u|\nu) \d x = \int \limits_{-K(\nu)}^{K(\nu)} \sqrt{u+\cnt(x|\nu)} \d x\quad.
\end{equation}
Equation~\eqref{eq:Lameabove} contains three unknown constants $\widetilde{D}_j$ which we need to determine from the properties of the action~\eqref{eq:SLameabove}. To do so, we use:
\begin{enumerate}
 \item the exact result for $\tilde{s}(u|\nu)$ at $u=\dfrac{1-\nu}{\nu}$,
 \item the fact that $\tilde{s}(u|\nu)$ is analytic near $u=\dfrac{1-\nu}{\nu}$, while the basis functions $G_1(u)$ and $G_2(u)$ are not,
 \item the logarithmic divergence of $\partial_u\tilde{s}(u|\nu)$ as $u\to0^{\s+}$.
\end{enumerate} %

From the first condition, it turns out that the integral in equation~\eqref{eq:SLameabove} can be evaluated analytically at $u_0=\dfrac{1-\nu}{\nu}$:
\begin{equation}
 \tilde{s}\biggl(\dfrac{1-\nu}{\nu}\biggr|\biggl.\nu\biggr) = \dfrac{\pi}{\nu} \quad.
\end{equation}
For convenience, we choose $u_0 = \dfrac{1-\nu}{\nu}$ to be the integration limit for the basis functions $G_1$ and $G_2$. Then both of them vanish at this point, ${G_{1,2}\biggl(\dfrac{1-\nu}{\nu},\dfrac{1-\nu}{\nu}\biggr|\biggl.\nu\biggr)=0}$, and the only remaining term is the constant~$\widetilde{D}_0$. Hence we obtain:
\begin{equation}\label{eq:D0above}
 \widetilde{D}_0
 = \dfrac{\pi}{\nu} \quad.
\end{equation}

Turning to the second condition, we can see that, physically, $u=\dfrac{1-\nu}{\nu}$ is not a special value for the energy. So the action is analytic in the vicinity of this point. However, the two basis functions are not, their non-analytic parts being
\begin{alignat}{9}
 g_1^{\text{\s n/a}}(u) &=& -\dfrac{\iu}{\pi}\sqrt{\dfrac{\nu}{1-\nu}}\log\left(u-\dfrac{1-\nu}{\nu}\right)\quad&&,\\
 g_2^{\text{\s n/a}}(u) &=& \dfrac{1}{2}\sqrt{\dfrac{\nu}{1-\nu}}\log\left(u-\dfrac{1-\nu}{\nu}\right)\quad&&.
\end{alignat}
\begin{sloppypar*}
Here we defined ${\displaystyle G_{1,2}(u,u_0|\nu)=\int \limits_{u_0}^u g_{1,2}(u)\d u}$, and took into account that $g_{1,2}^{\text{\s n/a}}(u)$ are the lowest-order non-analytic parts of the integrand. In order for these terms to cancel in equation~\eqref{eq:Lameabove}, we require
\end{sloppypar*}
\begin{equation}\label{eq:Lameaboveratio}
 \widetilde{D}_1 = \dfrac{\pi \iu}{2}\widetilde{D}_2 \quad.
\end{equation}

We now consider the third condition. As $u\to0^{\s+}$, the integrands of the basis functions in equation~\eqref{eq:solLame} diverge logarithmically. Along with equation~\eqref{eq:Lameaboveratio}, we get the following equality for the divergent part:
\begin{equation}
 \widetilde{D}_1\s g_1(u) + \widetilde{D}_2\s g_2(u) = \widetilde{D}_2\left(-\dfrac{\pi \iu}{2}g_1(u) + g_2(u)\right) \approx \widetilde{D}_2\dfrac{-\iu\log(u)}{2\sqrt{1-\nu}} + \mathcal{O}(1)\quad.
\end{equation}
At the same time, in the limit of $u\to0^{\s+}$, the derivative of the action becomes:
\begin{multline}
\partial_u\tilde{s}(u) = \int\limits_{-K(\nu)}^{K(\nu)} \dfrac{\d x}{2\sqrt{u+\cnt(x|\nu)}}
 \approx \int\limits_{K(\nu)-0}^{K(\nu)}\dfrac{\d x}{\sqrt{u+(1-\nu)(x-K(\nu)^2)}}
 \\\approx \dfrac{\log(u)}{2\sqrt{1-\nu}} + \mathcal{O}(1) \quad.
\end{multline}

Comparison of these two expressions yields: $\widetilde{D}_2=\iu$. Together with  expressions~\eqref{eq:D0above} and~\eqref{eq:Lameaboveratio}, this renders us the final result for the classical action above the maximum in the Lam\'e potential:
\begin{equation}
    \tilde{s}_c(u|\nu) = \dfrac{\pi}{\sqrt{\nu}} - \dfrac{\pi}{2} G_1 \biggl(\biggl.u,\dfrac{1-\nu}{\nu}\biggr|\nu\biggr) + \iu G_2 \biggl(\biggl.u,\dfrac{1-\nu}{\nu}\biggr|\nu\biggr)
    \quad.
\end{equation}

\section{Reconstructing the perturbative expansion from the quantum action}\label{app:recon}

We have already mentioned that the perturbative expansion can be generated by inverting the generalised Bohr-Sommerfeld quantisation condition~\eqref{GBS}. We now employ this observation to obtain an independent check of our results for $S(E)$ and~$\qactE_2(E)$.

\subsection{The self-dual sextic potential}
The GBS quantisation condition for the sextic potential~\eqref{eq:sexticpot} reads as:
\begin{equation}
    S(E) - \qactE_2(E) + \ldots = 2 \pi B
    \quad,\qquad\text{where}\qquad
    B = n+ \dfrac{1}{2} \quad.
\end{equation}
\begin{sloppypar*}
Here $S(E)$ is given by equation~\eqref{eq:Ssextic} with ${\widetilde{S}(u) = \widetilde{S}_1(u)}$, the function $\widetilde{S}_1(u)$ being defined by equations~\eqref{eq:actionsextic} and~(\ref{eq:sexticC13}-\ref{eq:sexticC11C12}). The function $\qactE_2(E)$ is given by equation~\eqref{sigma2final}. We choose the constants $b$ and $d$ to be:
\footnote{~The reasoning for our choice of constants $b$ and $d$ is the following: after changing variables $y=gx$, the potential acquires simple form ${\tilde{V}(y)=y^2(1+y^2)(3+4y+4y^2)/(6g^2) = (y^2+\ldots)/(6g^2)}$, for which an efficient method of constructing the perturbative expansion described in~\cite{UWKB} can be applied directly.}
\end{sloppypar*}
\begin{equation}
    b = \dfrac{1}{8}
    \quad,\qquad
    d = \dfrac{2}{3} g^4
    \quad.
\end{equation}
\begin{sloppypar*}
We then expand the expression on the LHS around ${E=V_{\text{min}} =
- \dfrac{1}{48g^2}}$, and invert the series term by term, in order to find $E(B,g)$. This renders:
\end{sloppypar*}
\begin{equation}
    E(N,g) = - \dfrac{1}{48g^2} + B
    - \left(\dfrac{5}{18}+\dfrac{20}{3}B^2\right)g^2
    - \left( \dfrac{100}{27}B + \dfrac{880}{27}B^3 \right) g^4 -\ldots\quad,
\end{equation}
which agrees with the regular perturbative expansion.

\subsection{The elliptic potential}

Following
\cite{Dual}, in the case of the elliptic potential~\eqref{eq:Ve}, we divide the Schr\"odinger equation~\eqref{Sch} by ${a=\kappa^2}$ and set $b = -\dfrac{\kappa^2}{2}$. This results in
\begin{equation}
    \biggl[-\dfrac{1}{\kappa^2}\dfrac{\d{}^2}{\d x^2} +
\nu \snt(x|\nu)\biggr] \psi(x) = \biggl[\dfrac{E}{\kappa^2}+\dfrac{1}{2}\biggr] \psi(x)
\quad.
\end{equation}
We now notice that here $\dfrac{1}{\kappa}$ is effectively playing the role of $\hbar$. With our equation cast into such a shape, the GBS quantisation condition for the lowest energy level takes the form of
\begin{equation}
    S(E_0|\nu) - \dfrac{1}{\kappa^2}\qactE_2(E_0|\nu) + \ldots = \dfrac{1}{\kappa}\pi
    \quad.
\end{equation}
Here $S(E_0|\nu)$ is given by equation~\eqref{eq:Slame}, with $s(u|\nu)=s_c(u|\nu)$ defined in equation~\eqref{eq:actionLame}; $\qactE_2(E_0|\nu)$ is given by~\eqref{sigmatla}. Being inverted term by term, this renders:
\begin{equation}
    E_0 = -\dfrac{1}{2}\kappa^2
    \left(
    1 - \dfrac{2\sqrt{\nu}}{\kappa}
    + \dfrac{\nu+1}{2\kappa^2}
    + \dfrac{1-4\s\nu+\nu^2}{8\sqrt{\nu}\s\kappa^3}
    +\ldots
    \right) \quad,
\end{equation}
which agrees with equation (70) in~\cite{Dual}.

\end{appendices}

\printbibliography[heading=bibintoc]

\end{document}